\documentclass[aps,prxquantum,reprint,superscriptaddress,nofootinbib,amsmath,amssymb,floatfix]{revtex4-2}
\usepackage{graphicx}
\usepackage{dcolumn}
\usepackage{bm}
\usepackage[utf8]{inputenc}
\usepackage{braket}
\usepackage{physics}
\usepackage{xcolor}
\usepackage[normalem]{ulem}
\usepackage[colorlinks=true,
            linkcolor=blue,
            citecolor=blue,
            urlcolor=blue,
            filecolor=blue,
            pdfborder={0 0 0}
           ]{hyperref}

\begin{document}
\preprint{APS/123-QED}

\title{Kinetic Cluster Embedding for Qubit Decoherence in Dissipative Spin Baths}

\author{Xiaofei Yu}
\altaffiliation{These authors contributed equally to this work}
\affiliation{Pritzker School of Molecular Engineering, The University of Chicago, Chicago, IL 60637}
\affiliation{Department of Physics, The University of Chicago, Chicago, IL 60637}
\author{Jonah Nagura}
\altaffiliation{These authors contributed equally to this work}
\affiliation{Pritzker School of Molecular Engineering, The University of Chicago, Chicago, IL 60637}
\author{Stella Wang}
\affiliation{Department of Physics, The University of Chicago, Chicago, IL 60637}
\author{Evan J. Villafranca}
\affiliation{Pritzker School of Molecular Engineering, The University of Chicago, Chicago, IL 60637}
\author{Peter C. Maurer}
\email{pmaurer@uchicago.edu}
\affiliation{Pritzker School of Molecular Engineering, The University of Chicago, Chicago, IL 60637}
\affiliation{Materials Science Division, Argonne National Laboratory, Lemont, IL, 60439, USA}
\affiliation{CZ Biohub Chicago, LLC, Chicago, IL 60642, USA}
\author{Giulia Galli}
\email{gagalli@uchicago.edu}
\affiliation{Pritzker School of Molecular Engineering, The University of Chicago, Chicago, IL 60637}
\affiliation{Materials Science Division, Argonne National Laboratory, Lemont, IL, 60439, USA}

\date{\today}

\begin{abstract}
Interpreting the decoherence of a spin qubit requires knowing how the surrounding bath spins behave: whether they fluctuate independently, exchange polarization incoherently, or evolve as coherent groups. These processes can yield the same Hahn-echo coherence time yet produce different coherence signals under pulse control of the qubit. Existing theoretical approaches either capture bath correlations at a cost that grows combinatorially with spin-bath cluster size or neglect polarization transport altogether. We introduce kinetic cluster embedding (KCE), which uses a single dimensionless parameter, comparing the flip-flop coupling of a spin pair with its detuning and linewidth, to decide how each pair must be treated. Pairs that can complete an appreciable exchange oscillation are propagated with the full dissipative quantum dynamics. All other spins are treated as random-telegraph fluctuators coupled through a kinetic network of polarization-transfer rates. We show that for nitrogen-vacancy centers coupled to surface electron spins, KCE propagates about 7 four-spin clusters, compared to more than 10$^5$ propagated by fourth-order master-equation cluster-correlation (ME-CCE4); nevertheless it reproduces ME-CCE4 coherence times to within 15$\%$  for dipolar couplings of order $10^{2}$ times the bath relaxation rate. For a three-dimensional P1 bath, the agreement is within 16$\%$.
Coherent multispin clusters emerge as the leading correction to independent fluctuators, and they leave a distinct signature in the CPMG response. Importantly, we find that whether a rate description suffices is set not by coupling strength alone but by whether a pair can exchange polarization coherently before detuning and damping suppress it.
\end{abstract}

\maketitle

\section{Introduction}
Understanding how the dynamics of an interacting open quantum system is encoded in a local probe is central to quantum sensing~\cite{degen2017quantum,sung2019non,davis2023probing,dwyer2022probing}. In many protocols, the dephasing of the probe (for example, a spin qubit) is used to infer the correlations, transport, and relaxation timescales of the bath surrounding it~\cite{Zu2021,Jackson2021,Gangloff2021,Rezaiprobing2025}. Dephasing measurements are commonly interpreted by representing the bath's effect as a stochastic, time-dependent field acting on the probe, characterized by a noise spectrum and a correlation time~\cite{PhysRevLett.120.243604,PhysRevResearch.7.013333,witzel2012quantum,Zu2021}. The microscopic information extracted from such measurements is reliable only if the assumed bath model accounts for every microscopic process that causes fluctuations in the bath-induced field~\cite{Yang2008,Witzel2014,RevModPhys.82.1155,Szankowski2018,Onizhuk2025}.

In a dissipative spin bath, fluctuations of the field acting on the probe can arise from intrabath flip-flops as well as relaxation processes within the bath. Flip-flops coherently couple bath-spin pairs at a rate set by the dipolar coupling, while bath-spin relaxation and pure dephasing cause the resulting pair coherence to decay, giving it a finite linewidth. Although physically distinct, flip-flops and relaxation can contribute comparably to probe dephasing, so the measured coherence time $T_2$ alone cannot identify the dominant mechanism~\cite{PhysRevLett.120.243604,PhysRevResearch.7.013333,witzel2012quantum,Zu2021}.  Understanding the different processes causing decoherence is especially important for solid-state qubits limited by dipolar-coupled electron-spin baths, including substitutional nitrogen in diamond and paramagnetic defects at surfaces and interfaces~\cite{de2021materials,dwyer2022probing,joos2022protecting,Denis_C,nagura2025}. More broadly, identifying when polarization exchange requires a coherent description and when it admits an effective kinetic one connects this problem to dipolar quantum simulators based on polar molecules, Rydberg atoms, and dense color-center ensembles~\cite{yan2013observation,Kucsko2018,glassy2021, christakis2023probing, bluvstein2019extending}.

We focus on shallow nitrogen-vacancy (NV) centers coupled to a two-dimensional (2D) bath of paramagnetic surface spins, whose fluctuations limit the coherence available for nanoscale sensing~\cite{degen2017quantum,de2021materials,dwyer2022probing}. This geometry provides a useful control parameter: the ratio of NV depth $d$ to mean surface-spin spacing $d_{\mathrm{ee}}$ varies the probe's sensitivity from a few strongly coupled spins at $d\lesssim d_{\mathrm{ee}}$ to many weaker contributions at $d\gg d_{\mathrm{ee}}$~\cite{dwyer2022probing,nagura2025}. Identifying the underlying dynamics is therefore necessary both to interpret bath parameters inferred from NV coherence and to predict the effects of bath driving or surface modification~\cite{dwyer2022probing,bluvstein2019extending,sangtawesin2019origins}.

Master-equation cluster-correlation expansion methods (ME-CCE) describe coherent and dissipative dynamics within chosen spin clusters, organizing correlations by cluster cardinality~\cite{Yang2008,onizhuk2024understanding,Onizhuk2025}. They consequently treat at the same order pairs whose coherent flip-flops are suppressed by detuning or dissipation and near-resonant pairs that exchange polarization coherently before losing coherence. Larger correlated groups require higher-order clusters at combinatorially increasing cost, without an explicit separation of coherent exchange from incoherent polarization transfer~\cite{Yang2008,schaetzle2024spin,Kucsko2018}. Independent-fluctuator models instead represent each bath spin as a local two-state field source, neglecting correlations generated by polarization transport~\cite{Klauder1962,deSousa2003,Ajoy2012,Dikarov2016,Jurcevic_2022}. Their validity requires polarization transfer to be slow compared with intrinsic relaxation; successive flip-flops across several spins or few coherent exchange within near-resonant pairs can invalidate this approximation.  
Neither approach by itself supplies a microscopic criterion for separating decoherence dominated by independent relaxation, coherent pairs, or collective polarization transport within a cluster of spins. This distinction is needed to predict the response to pulse control, the depth and density dependence of coherence decay times and shapes, and whether rate equations or explicit many-body propagation are required for a quantitative prediction of coherence times.

Here we develop a microscopic theoretical framework, called kinetic cluster embedding (KCE), that identifies the nature of the decoherence mechanism directly from the bath Hamiltonian and its dissipative rates. We show that whether polarization is exchanged coherently or incoherently within a pair of bath spins depends on a single dimensionless parameter defined in terms of its dipolar coupling, detuning, and linewidth. Incoherent pairs form a kinetic network, in which polarization hops stochastically between spins according to a set of rate equations;  while coherent pairs are grouped into explicit clusters propagated with the full dissipative quantum dynamics. 
 The criterion is simple to evaluate, provides a microscopic interpretation of probe decoherence, and offers a route to extending KCE beyond surface baths to systems with the same exchange and local-dissipation structure.

\section{Exact local building block} 
We first consider one bath spin $e$ longitudinally coupled to a probe spin $v$ with strength $A$. In the toggling frame of the chosen control sequence, the
Hamiltonian is given by:
\begin{equation}
  \hat H_N(t)
  =\frac{A}{4}\,y_N(t)\,
  \hat\sigma_z^{(v)}\hat\sigma_z^{(e)},
  \label{eq:toggling_hamiltonian}
\end{equation}
where $y_N(t)=\pm1$ changes sign at each of the $N$ ideal probe $\pi$ pulses. Throughout, couplings are expressed in frequency units as $A/(2\pi)$. Dissipation is characterized by the longitudinal-relaxation
and pure-dephasing rates $\gamma_{v1}$ and $\gamma_{v2}$ of the
probe spin, and by the corresponding rates $\gamma_{e1}$ and
$\gamma_{e2}$ of the bath spin.
The normalized coherence of the probe is $\mathcal L_N(t)=2\operatorname{Re}\operatorname{Tr}_e\hat\tau_N(t)$, where $\hat\tau_N(t)=\bra{0_v}\hat\rho_N(t)\ket{1_v}$ is the off-diagonal block of the joint density operator $\hat\rho_N$. In the bath-spin Pauli basis \cite{nielsen2000quantum}, this block is given by:
\begin{equation}
\begin{aligned}
\hat{\tau}_N(t)
={}&\frac{1}{4}\Bigl[
u_N(t)\,\hat{\sigma}_0^{(e)}
+m_N(t)\,\hat{\sigma}_z^{(e)}
\\
&\quad
+p_N(t)\,\hat{\sigma}_x^{(e)}
+q_N(t)\,\hat{\sigma}_y^{(e)}
\Bigr].
\end{aligned}
\end{equation}
 Taking the trace of this expansion yields $\mathcal L_N(t)=\operatorname{Re}[u_N(t)]$. Substituting this expansion into the Lindblad master equation gives one equation of motion for each of the four coefficients $u$, $m$, $p$, and $q$ (see Appendix~\ref{Noneequalrate}). Under longitudinal coupling, $u_N$ and $m_N$ form a closed pair of equations, while $p_N$ and $q_N$ are decoupled. The equations for $u_N$ and $m_N$ determine $\mathcal L_N(t)$ and explicitly contain the central spin's intrinsic coherence-decay rate $\Gamma_v=\gamma_{v1}/2+\gamma_{v2}$. Writing $u_N(t)=e^{-\Gamma_vt}\widetilde u_N(t)$ and $m_N(t)=e^{-\Gamma_vt}\widetilde m_N(t)$, the dependence on $\Gamma_v$ can be eliminated leaving the closed, $\Gamma_v$-independent linear system:

\begin{equation}
  \frac{d}{dt}
  \begin{pmatrix}
    \widetilde u_N\\[1pt]
    \widetilde m_N
  \end{pmatrix}
  =
  \begin{pmatrix}
    0 & -iAy_N(t)/2\\
    -iAy_N(t)/2 & -\gamma_{e1}
  \end{pmatrix}
  \begin{pmatrix}
    \widetilde u_N\\[1pt]
    \widetilde m_N
  \end{pmatrix}.
  \label{eq:reduced_2x2_system}
\end{equation}

Equation~\eqref{eq:reduced_2x2_system} shows that the probe-bath coupling mixes $\widetilde u_N$ and $\widetilde m_N$, while longitudinal relaxation of the bath spin damps $\widetilde m_N$ at rate $\gamma_{e1}$. Bath pure dephasing, $\gamma_{e2}$, does not enter these equations and therefore does not affect the probe coherence.

The probe coherence is $\mathcal L_N(t)=e^{-\Gamma_vt}\operatorname{Re}\tilde u_N(t)$. Equation~\eqref{eq:reduced_2x2_system} is also the conditional-coherence equation for a classical random-telegraph field $\xi(t)=\pm A/2$ switching at the rate $\gamma_{e1}/2$ \cite{bergli2007exact,bergli2009decoherence}; thus, for any probe pulse sequence, the relaxing bath spin produces the same probe coherence as this classical fluctuator, even if the bath spin initially has transverse coherence. In the fast-switching limit $|A|\ll\gamma_{e1}$, the field changes sign before the probe accumulates appreciable phase, and the long-time Ramsey envelope approaches $\widetilde{\mathcal L}_0(t)\simeq\exp[-A^2t/(4\gamma_{e1})]$ to leading order. The random-telegraph mapping remains exact outside this limit, but the probe coherence then depends on the full  switching statistics encoded in Eq.~\eqref{eq:reduced_2x2_system}, rather than only on the noise spectrum.  Solutions corresponding to the Ramsey, Hahn-echo, and CPMG sequences are given in Appendix~\ref{Noneequalrate} and \ref{connectiontospectrum}. 

We now use Eq.~\eqref{eq:reduced_2x2_system} as the exact single-spin building block to determine which bath-bath couplings require coherent propagation and which admit a polarization-transfer rate.

\section{Interacting baths: from coherent flip-flops to stochastic transfer}
We consider bath spins that interact through their dipolar coupling. Its secular part is
\begin{equation}
\hat H_B=
\sum_{i<j}J_{ij}\left(
\hat\sigma_+^{(i)}\hat\sigma_-^{(j)}+
\hat\sigma_-^{(i)}\hat\sigma_+^{(j)}\right)
+\sum_{i<j}U_{ij}\hat\sigma_z^{(i)}\hat\sigma_z^{(j)},
\label{eq:bath_hamiltonian}
\end{equation}
where $\hat\sigma_\pm^{(i)}$ and $\hat\sigma_z^{(i)}$ are the raising, lowering, and Pauli $z$ operators of bath spin $i$. The dipolar flip-flop coupling $J_{ij}$ exchanges polarization between two spins, and the longitudinal coupling $U_{ij}=-J_{ij}$ shifts their energies (Appendix~\ref{meanfieldapproximation}).
\subsection{Pair reduction and coherent clusters}
Consider a bath-spin pair $(i, j)$ with probe couplings $A_i$ and $A_j$ and flip-flop coupling $J_{ij}$. We treat the remaining bath spins as spectators with fixed $z$ polarizations $s_k=\pm1$. The flip-flop term of Eq.~\eqref{eq:bath_hamiltonian} conserves the pair's total $z$ projection and couples the two zero-projection states, $\ket{\uparrow_i\downarrow_j}$ and $\ket{\downarrow_i\uparrow_j}$. For probe state $\ket a$, with $a\in \{0,1\}$, and spectator configuration $\mathbf s=(s_k)_{k\ne i,j}$, the pair Hamiltonian is~\cite{yao2006}

\begin{equation}
\begin{aligned}
\hat H_{ij}^{(a)}&=J_{ij}\hat X+
\frac{\Delta_{ij}^{(a)}(\mathbf s)}{2}\hat Z,\\
\Delta_{ij}^{(a)}(\mathbf s)&={\nu}_a(A_i-A_j)
+2\!\sum_{k\ne i,j}(U_{ik}-U_{jk})s_k .
\end{aligned}
\label{eq:pair_hamiltonian_main}
\end{equation}

Here $\hat X$ and $\hat Z$ are Pauli operators in the two-state subspace and $\nu_a=\langle a|\hat S_z^{(v)}|a\rangle$ is the probe spin projection in state $\ket a$. Together, the probe-coupling difference $(A_i-A_j)$ and the spectator shifts set the pair detuning.

For uniform bath-spin rates, longitudinal relaxation and pure dephasing damp the pair coherence at $\Gamma_{ij}=\gamma_{e1}+2\gamma_{e2}$, while the population difference of the pair decays at $\gamma_{e1}$. Because these two rates are comparable, the pair coherence cannot in general be eliminated adiabatically (Appendix~\ref{sec:pair_projection_sm}). For a fixed probe branch $a$ and spectator configuration $\mathbf s$, we instead replace the pair by the transfer rate that reproduces its time-integrated polarization difference exactly, for any $J_{ij}$, and use its average,
\begin{equation}
\begin{aligned}
W_{ij}^{(a)}(\mathbf s)&=
\frac{2J_{ij}^2\Gamma_{ij}}
{[\Delta_{ij}^{(a)}(\mathbf s)]^2+\Gamma_{ij}^2},\\
\overline W_{ij}&=\sum_a p_a
\left\langle W_{ij}^{(a)}(\mathbf s)\right\rangle_{\mathbf s}.
\end{aligned}
\label{eq:pair_rate_main}
\end{equation}
Here $p_a$ is the probe population in state $\ket a$, and $\langle\cdots\rangle_{\mathbf s}$ denotes an average over spectator configurations. The averaged rate $\overline W_{ij}$ transfers
polarization between the spins while conserving their total polarization. When $|J_{ij}|\ll\sqrt{[\Delta_{ij}^{(a)}(\mathbf s)]^2+\Gamma_{ij}^2}$, the exchange of the pair is weak within its lifetime and the rate description reproduces the pair dynamics up to an error of order $\lambda_{ij}^2$, with $\lambda_{ij}$ defined below, comparable to the exchange contribution itself; it becomes a Markovian rate only when bath dephasing dominates, $\gamma_{e2}\gg\gamma_{e1}$. Its use for probe coherence is a further approximation: the same reduction carried out in the probe's off-diagonal block evaluates the rate at the mean of $\Delta_{ij}^{(0)}(\mathbf s)$ and $\Delta_{ij}^{(1)}(\mathbf s)$
 (Appendix~\ref{sec:pair_projection_sm}); all results below use $\overline W_{ij}$.
To assess the pair-rate approximation, we define the dimensionless parameter
\begin{equation}
\lambda_{ij}^2=
\sum_a p_a\left\langle
\frac{J_{ij}^2}{[\Delta_{ij}^{(a)}(\mathbf s)]^2+\Gamma_{ij}^2}
\right\rangle_{\mathbf s}
=\frac{\overline W_{ij}}{2\Gamma_{ij}}.
\label{eq}
\end{equation}
When $\lambda_{ij}\ll1$, exchange is weak relative to detuning and damping on average, supporting a rate description. Pairs with $\lambda_{ij}$ of order unity require explicit coherent propagation. Because $\lambda_{ij}$ is averaged over the two probe states and spectator configurations, a small value does not guarantee the weak-pair condition is satisfied in every configuration.

We define a cutoff $\lambda_c$. Spins belong to the same cluster
if they are connected by a chain of pairs satisfying
$\lambda_{ij}>\lambda_c$. This rule partitions the bath into
disjoint clusters $\mathcal C_\alpha$. A spin with no
above-threshold partner forms a cluster by itself, which we call a
singleton. Within each multispin cluster, we propagate the full
dissipative quantum dynamics, including internal couplings below
the cutoff. Since the Liouvillian dimension grows as $4^k$ for a $k$-spin cluster, we retain at most four spins per cluster. Every pair connecting distinct clusters, including above-cutoff pairs excluded by this size limit, is represented by the polarization-transfer rate $\overline W_{ij}$ (Appendix~\ref{sec:clusterCountsSM}).

\subsection{Kinetic network of polarization transfer}
The longitudinal polarizations $m_i(t)=\langle\hat\sigma_z^{(i)}(t)\rangle$
track how rate-treated flip-flops redistribute polarization through the bath. When all flip-flop couplings are reduced to rates, their
evolution under intrinsic relaxation and polarization transfer is
\begin{equation}
\dot m_i=-\gamma_{e1}m_i
-\sum_{j\ne i}\overline W_{ij}(m_i-m_j).
\label{eq:kinetic_network_main}
\end{equation}
The spins connected by the rates $\overline W_{ij}$ form the kinetic network. Each rate describes a joint transfer between spins with opposite $z$ projections. Their contributions to $\dot m_i$ and $\dot m_j$ have opposite signs, so exchange conserves their combined polarization, unlike intrinsic relaxation $\gamma_{e1}$. Successive transfers can return polarization to a spin or carry it through the network. When quantum clusters are retained, only intercluster flip-flop couplings enter the rate term; intracluster dynamics require a  quantum mechanical description.

\begin{figure*}[!t]
\centering
\includegraphics[width=0.75\textwidth]{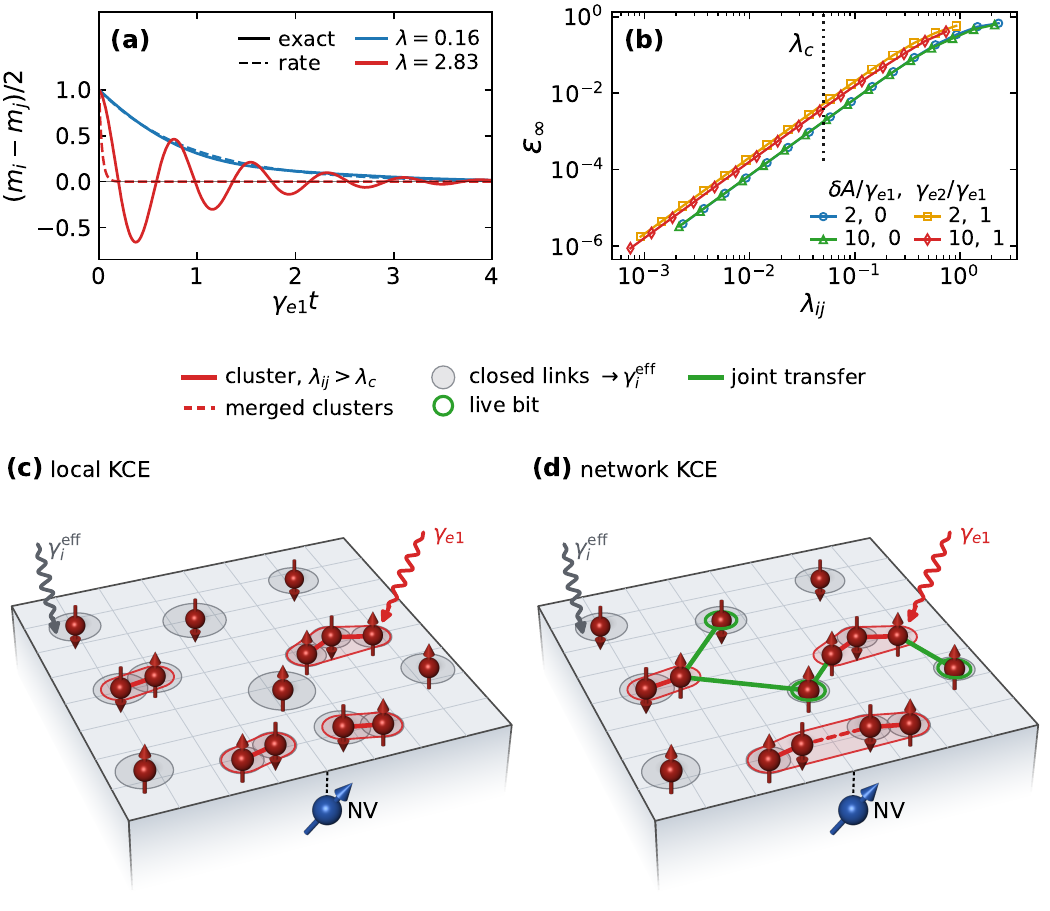}
\caption{Construction of KCE.
(a) Exact (solid) and rate-equation (dashed) pair polarization difference for $\lambda_{ij}=0.16$ and $2.83$; the rate misses the coherent oscillations of the strong pair.
(b) Spin-echo error $\epsilon_\infty$ of the pair-rate approximation versus $\lambda_{ij}$ for several $\delta A/\gamma_{e1}$ ($\delta A=A_i-A_j$) and $\gamma_{e2}/\gamma_{e1}$; dotted line: illustrative $\lambda_c$.
(c,d) Schematic of the partition (illustrative bath, not to scale): bath spins on the diamond surface, with the NV (blue) in the cut face below. Red pulses: intrinsic relaxation $\gamma_{e1}$ of the bath spins; gray pulses: rates $\gamma_i^{\rm eff}$ from closed links. (c) Local KCE: spins joined by $\lambda_{ij}>\lambda_c$ (red) form explicit clusters of at most four spins (outlined) that evolve independently. Every other link is closed into the rate $\gamma_i^{\rm eff}$ of both partners (gray discs, size growing with the number of closed links), so different units share no correlations. (d) Network KCE restores selected correlations: links to live bits (green rings) are sampled as joint transfers in which both partners flip together (green), and a link between two clusters merges them into one coherent group (dashed red). Only the remaining links are closed into $\gamma_i^{\rm eff}$, so the discs shrink.}
\label{fig:construction}
\end{figure*}

\subsection{Network-resolved dynamics}\label{sec:hsle}

To compute probe coherence, we represent the rate-treated 
intercluster pairs by stochastic exchanges: when its spins have
opposite $z$ projections, they exchange states at rate
$\overline W_{ij}$. The mean polarizations of these joint events are described by
Eq.~\eqref{eq:kinetic_network_main}. The
multispin clusters evolve quantum mechanically.

Let $\mathcal N$ denote the singleton spins, and let
$\mathbf s=(s_k)_{k\in\mathcal N}$, with $s_k=\pm1$, specify their
configuration. Intrinsic flips and exchanges between singleton spins
form a classical Markov jump process with generator $K$. The
probe-coherence block conditioned on $\mathbf s$,

\begin{equation}
\hat\tau_{\mathbf s}(t)
=\langle 0_v,\mathbf s|\hat\rho(t)|1_v,\mathbf s\rangle,
\end{equation}

is an operator on the multispin clusters. The coupled evolution
with the jump process is described by a hybrid stochastic Liouville equation
(hSLE)~\cite{gardiner2004quantum,Liouville1, Diosi, Nicola_2020}:

\begin{equation}
\partial_t\hat\tau_{\mathbf s}
=\mathcal K_{\mathcal C}[\mathbf s,t]\hat\tau_{\mathbf s}
-i y_N(t)\Phi(\mathbf s)\hat\tau_{\mathbf s}
+\sum_{\mathbf s'}K_{\mathbf s\mathbf s'}\hat\tau_{\mathbf s'}
+\mathcal T_{\mathbf s},
\label{eq:hsle_main}
\end{equation}

Here $\mathcal K_{\mathcal C}$ propagates the clusters conditioned on
$\mathbf s$ and includes the probe's intrinsic decay, while
$\Phi(\mathbf s)=(\nu_0-\nu_1)\sum_{k\in\mathcal N}A_ks_k/2$ is the
field of the singleton spins. The coefficients $K_{\mathbf s\mathbf s'}$ describe the classical
Markov dynamics of singleton flips and exchanges. The term $\mathcal T_{\mathbf s}$
accounts for joint exchanges involving cluster spins: it affects
both participating spins and, when one spin is a singleton, couples blocks
with different configurations. Its explicit form and numerical treatment are given in Appendix~\ref{app:hsle}. The probe coherence follows by summing over
configurations and tracing out the clusters:

\begin{equation}
\mathcal L_{\mathrm{KCE}}^{\mathrm{net}}(t)
=\operatorname{Re}\!\left[
\frac{\sum_{\mathbf s}\operatorname{Tr}\hat\tau_{\mathbf s}(t)}
{\sum_{\mathbf s}\operatorname{Tr}\hat\tau_{\mathbf s}(0)}
\right].
\label{eq:hsle_coherence_main}
\end{equation}

\subsection{Local closure of the kinetic network}\label{sec:closure}
We approximate the coupled dynamics of the kinetic network by
treating each intercluster coupling with an associated rate as a separate
contribution to local spin relaxation. For $i\in\mathcal C_\alpha$,
this corresponds to setting the external spin's mean polarization
$m_j\simeq0$ in Eq.~\eqref{eq:kinetic_network_main}, yielding:

\begin{equation}
\gamma_i^{\mathrm{eff}}
=\gamma_{e1}
+\sum_{j\notin\mathcal C_\alpha}\overline W_{ij}.
\label{eq:gammaeff_main}
\end{equation}

For a multispin cluster, the sum contains only couplings to spins
outside the cluster; its internal dynamics is quantum mechanical.
We call this formulation the local closure of the kinetic network:
intercluster couplings enter the rates $\gamma_i^{\mathrm{eff}}$
instead of generating joint transfers, and the clusters are
propagated independently.
The closure applies at any $\lambda_c$. In the limit
$\lambda_c\to\infty$, every cluster is a singleton and all exchange
contributions enter the local rates. The closure discards
correlations from joint intercluster transfers, including
polarization return and transport through successive pairs. 
Because the clusters are propagated independently in the local
description, their bath-induced coherence amplitudes can be factorized\cite{disjoint}:

\begin{equation}
\mathcal L_{\mathrm{KCE}}^{\mathrm{loc}}(t)
=e^{-\Gamma_vt}\operatorname{Re}
\left[\prod_\alpha
\widetilde\ell_{\mathcal C_\alpha}(t)\right].
\label{eq:Lhybrid_main}
\end{equation}

Here $\widetilde\ell_{\mathcal C_\alpha}$ is the complex coherence
amplitude of cluster $\mathcal C_\alpha$ with the probe's intrinsic
decay removed. For a singleton, we use 
Eq.~\eqref{eq:reduced_2x2_system} with
$\gamma_{e1}$ replaced by $\gamma_i^{\mathrm{eff}}$.
Each multispin factor is obtained by propagating its cluster
master equation with the corresponding local rates. In the numerical implementation the two formulations are connected by
a second threshold $\lambda_w\le\lambda_c$: rate-treated links with
$\lambda_w<\lambda_{ij}\le\lambda_c$ are sampled as joint events, while
those with $\lambda_{ij}\le\lambda_w$ enter the local rates
$\gamma_i^{\mathrm{eff}}$, so that $\lambda_w=\lambda_c$ recovers the
local closure and $\lambda_w=0$ retains every sampled transfer. All
network-KCE results use $\lambda_w=0.1$, for which the sampled
coherence is converged, and evaluate
Eq.~\eqref{eq:hsle_coherence_main} with the trajectory estimator of
Eq.~\eqref{eq:net_KCE} (Appendix~\ref{app:hsle}).


{Figure~\ref{fig:construction} summarizes the construction of the kinetic network. Panel (a) compares coherent pair dynamics with its rate approximation, panel (b) shows that the absolute spin-echo error of this approximation falls as $\lambda_{ij}^2$ for weak pairs, guiding
the choice of $\lambda_c$, and panel (c) illustrates local KCE, where the threshold $\lambda_c$ determines the quantum clusters, while intercluster exchanges are incorporated into effective local relaxation rates . The singletons are treated as independent random-telegraph fluctuators, and the quantum clusters are propagated independently. Panel (d) illustrates network KCE, where selected intercluster exchanges are retained as stochastic joint transfers. The participating singletons are tracked as live bits, whose configurations condition the quantum-cluster dynamics and whose fluctuating fields contribute to the accumulated probe phase.

\begin{figure}[!t]
\centering
\includegraphics[width=0.8\columnwidth]{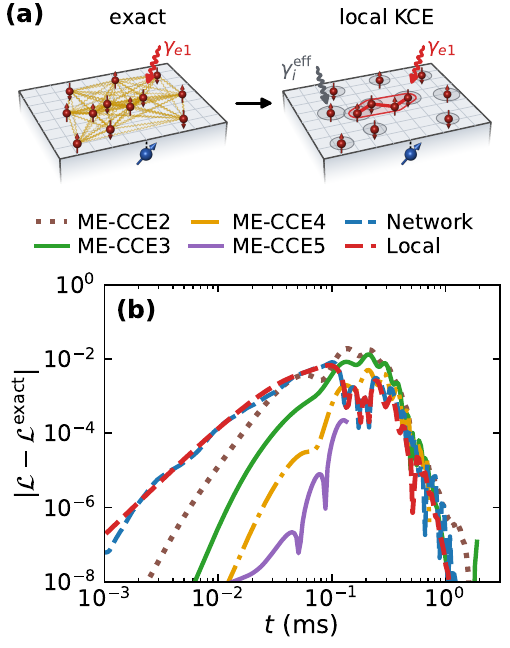}
\caption{Exact benchmark for a 12-spin bath. (a)~Schematic (illustrative positions): the NV (blue) coupled to an interacting dissipative bath in which every dipolar coupling is coherent (gold, left), and its local KCE representation (right). The nonperturbative group is propagated explicitly as a cluster (red, outlined), while the rate-controlled couplings are closed into independent relaxation rates $\gamma_i^{\rm eff}$ of the bath spins (gray discs). Red pulses: intrinsic relaxation $\gamma_{e1}$, present in both descriptions; gray pulse: flip-flop-induced rates $\gamma_i^{\rm eff}$. (b)  Absolute pointwise Hahn-echo error
$|\mathcal L(t)-\mathcal L^{\mathrm{exact}}(t)|$ relative to full
Liouvillian propagation. Results for ME-CCE2--5 and for network and local KCE, at $\rho=0.001~\mathrm{nm}^{-2}$, $d=10$~nm,
$T_{1e}=80~\mu\mathrm{s}$, for both KCE calculations. KCE formulations reach ME-CCE4 accuracy while propagating only single spins and selected multispin clusters}
\label{fig:validation}
\end{figure}

\section{Results}
\subsection{Benchmark and range of validity}
\label{bench}
We first benchmark network-resolved KCE and its local closure at
the same cutoff, $\lambda_c=0.5$, against exact full-Liouvillian
propagation for the 12-spin dissipative bath in
Fig.~\ref{fig:validation}. For each calculation $X$, we use the
maximum pointwise coherence error,
$\epsilon_\infty^{(X)}
=\max_t|\mathcal L_X(t)-\mathcal L^{\mathrm{exact}}(t)|$.
The comparison of the two calculations illustrates the effect of retaining joint
polarization transfers between distinct clusters.
Network-resolved KCE and its local closure yield
$\epsilon_\infty^{(\mathrm{net})}=6.0\times10^{-3}$ and
$\epsilon_\infty^{(\mathrm{loc})}=6.3\times10^{-3}$, respectively,
at the level of ME-CCE4 ($5.0\times10^{-3}$). The KCE error is
smaller than those of ME-CCE2 and ME-CCE3 and comparable to that
of ME-CCE4, where ME-CCE($k$) denotes truncation at $k$-spin clusters. 

\begin{figure*}[!t]
\centering
\includegraphics[width=0.85\textwidth]{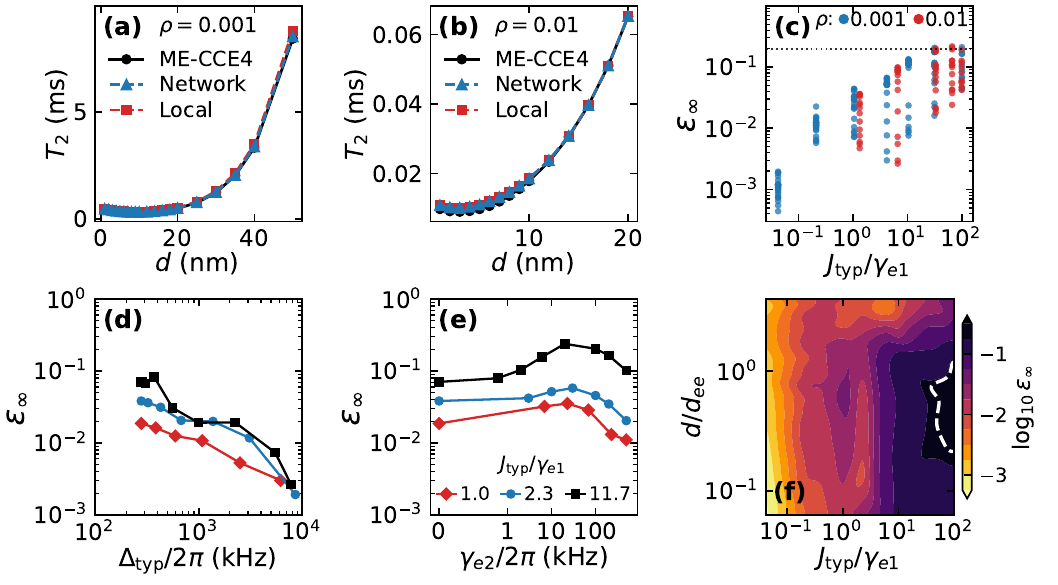}
\caption{Accuracy of network-resolved KCE and its local closure.
(a,b)~Hahn-echo $T_2$ versus NV depth at $T_{1e}=20~\mu$s for
(a)~$\rho=0.001$ and (b)~$0.01$~nm$^{-2}$: ME-CCE4 (black circles), network KCE (blue triangles) and local closure (red squares), both at $\lambda_c=0.5$. (c)~Maximum pointwise error $\epsilon_\infty$ of network KCE relative to ME-CCE4 versus $J_{\mathrm{typ}}/\gamma_{e1}$ at $\rho=0.001$ and $0.01$~nm$^{-2}$.(d,e)~$\epsilon_\infty$ at $\rho=0.005$~nm$^{-2}$ and $d=10$~nm for three values of $J_{\mathrm{typ}}/\gamma_{e1}$, versus (d)~the typical pair detuning $\Delta_{\mathrm{typ}}$, raised by on-site disorder of width $\sigma_\varepsilon$ ($\gamma_{e2}=0$), and (e)~the bath dephasing rate $\gamma_{e2}$ ($\sigma_\varepsilon=0$). (f)~$\epsilon_\infty$ interpolated over all 280 parameter sets in the $(J_{\mathrm{typ}}/\gamma_{e1},\,d/d_{ee})$ plane. The dotted line in (c) and the dashed contour in (f) mark $\epsilon_\infty=0.2$. }
\label{fig:validitymap}
\end{figure*}

We next compare the two KCE formulations for a negatively charged NV center at depth $d$ below a $(100)$ diamond surface, coupled to paramagnetic surface defects modeled as a two-dimensional bath of electron spins. We consider areal densities
$\rho=0.001$, $0.005$, and $0.01$~nm$^{-2}$ and bath relaxation
times $T_{1e}$ spanning representative values reported or inferred
for diamond surfaces~\cite{davis2023probing,yu2025engineering,zheng2022coherence,de2018density,lo2025enhancement}.{
We compare the network-resolved KCE and its local closure against ME-CCE4~\cite{yang2016quantum,witzel2005quantum,Onizhuk2025,onizhuk2024understanding}, averaging each data point over 100 bath realizations. Convergence tests show that ME-CCE results are converged at fourth order (see Appendix \ref{benchmarks})}. {For a random spin bath on a surface of areal density $\rho$, the mean nearest-neighbor distance is $d_{\mathrm{ee}}=1/(2\sqrt{\rho})$~\cite{dwyer2022probing, yu2025engineering}. We define the characteristic flip-flop coupling $J_{\mathrm{typ}}=\frac{2}{\pi}\,\frac{\mu_0\gamma_e^{2}\hbar}{16\pi\,d_{\mathrm{ee}}^{3}}
=\frac{\mu_0\gamma_e^{2}\hbar}{\pi^{2}}\,\rho^{3/2}$  as the dipolar flip-flop matrix element of a pair at this separation, averaged over the in-plane orientation of the pair, i.e. $J_{\mathrm{typ}}/2\pi=66~\mathrm{kHz}\,(\rho/0.01~\mathrm{nm}^{-2})^{3/2}$.} Under Hahn echo, we examine the NV depth dependence from $d=1$~nm to beyond $3\,d_{ee}$ at bath relaxation times $T_{1e}$ ranging from $3.2~\mu\mathrm{s}$ to $1.6$~ms.

Figure~\ref{fig:validitymap}(a,b) compares network-resolved KCE
with its local closure at the same $\lambda_c=0.5$ to determine when
correlations between rate-treated intercluster transfers affect the
NV coherence as bath density and NV depth vary. At $\rho=0.001~\mathrm{nm}^{-2}$ and $T_{1e}=20~\mu\mathrm{s}$, $J_{\mathrm{typ}}/\gamma_{e1}\simeq0.25$: the typical flip-flop coupling is well below the relaxation rate, and the bath is relaxation dominated. The KCE partition of this bath, which contains about 200 spins, consists on average of 123 singletons and 36 two-spin clusters per realization, with three- and four-spin clusters occurring less than once per realization (Appendix~\ref{tab:clusterCountsSM}). 
 Across the depth sweep, the two formulations predict similar $T_2$, while network-resolved KCE remains close to ME-CCE4, with $\epsilon_\infty\leq0.016$.
Correlations between rate-treated transfers therefore have little effect on $T_2$ in this dilute bath. 
At $\rho=0.01~\mathrm{nm}^{-2}$ and $T_{1e}=20~\mu\mathrm{s}$,
$J_{\mathrm{typ}}/\gamma_{e1}\simeq8.3$: the typical flip-flop coupling exceeds the relaxation rate and the intrabath flip-flops dominate. The KCE partition of this bath, which contains about 200 spins, consists on average of 40 singletons, 51 two-spin, 9 three-spin and 7 four-spin clusters per realization, so that most spins belong to explicitly propagated clusters (Appendix~\ref{sec:clusterCountsSM}). For the same bath realizations, the ME-CCE4 connectivity graph contains roughly $5.5\times10^5$ connected
four-spin clusters that required explicit quantum propagation. Figure~\ref{fig:validitymap}(b) shows that both network KCE and its local closure similarly follow ME-CCE4 in $T_2$ over the full depth range, although the pointwise error exceeds $0.03$ for $d\lesssim2\,d_{ee}$. Thus the 107 clusters of the KCE partition, 67 of which contain two to four spins, capture the coherence time that ME-CCE4 recovers only by propagating more than $10^{5}$ four-spin clusters, and retaining the joint intercluster transfers in network KCE does not change the result in this regime. At slower relaxation, network KCE continues to follow the ME-CCE4 coherence time, whereas the local closure deviates because it discards the return of transferred polarization between clusters (Appendix~\ref{benchmarks}).

\begin{figure*}[!t]
\centering
\includegraphics[width=0.75\textwidth]{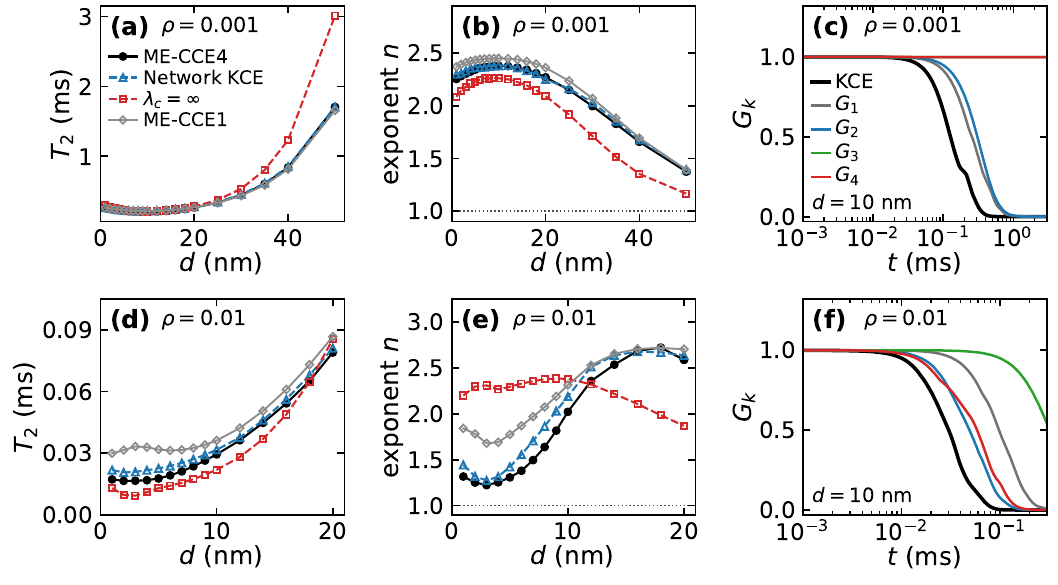}
\caption{Exchange, relaxation and cluster structure in the echo decay at $T_{1e}=100~\mu$s for $\rho=0.001$ (top row) and $0.01~\mathrm{nm}^{-2}$ (bottom row), i.e.\ $J_{\mathrm{typ}}/\gamma_{e1}\simeq1.3$ and $42$. (a,d) Hahn-echo $T_2$ and (b,e) stretching exponent $n$ versus NV depth averaged over 100 bath realizations, computed with ME-CCE4 (black), network KCE at $\lambda_c=0.5$ (blue), the singleton-local limit $\lambda_c\to\infty$ (red), where pair coupling enters only through effective switching rates, and ME-CCE1 (grey), where flip-flops are neglected. (c,f) Cluster-size decomposition of local KCE at $\lambda_c=0.5$ and $d=10$~nm. Each spin belongs to exactly one cluster, and $G_k(t)=\prod_{|\mathcal C_\alpha|=k}\widetilde\ell_{\mathcal C_\alpha}(t)$ is the product of the coherence factors of all $k$-spin clusters; black shows the full response, $\prod_k G_k$. In the dilute bath singletons and pairs carry the decay; in the dense bath pairs and four-spin clusters dominate.}
\label{fig:purelocal}
\end{figure*}

Figures~\ref{fig:validitymap}(c,f) map the agreement between network-resolved KCE and ME-CCE4 across 280 parameter sets spanning the three densities, $J_{\mathrm{typ}}/\gamma_{e1}$ from 0.04 to 100, and NV depths up to about $4\,d_{ee}$ . Over all of them, network KCE give $T_2$ within 22\% of ME-CCE4 for $J_{\mathrm{typ}}/\gamma_{e1}\lesssim 100$ at every depth. We quantify the pointwise deviation by $\epsilon_\infty^{(\mathrm{net})}=\max_t|\mathcal L^{\mathrm{net}}_{\mathrm{KCE}}(t)-\mathcal L_{\mathrm{ME\text{-}CCE4}}(t)|$. For each bath realization, KCE gives the coherence averaged over bath states directly, whereas ME-CCE4 samples them. The ME-CCE4 sampling error ranges from $10^{-3}$ in dilute, fast-relaxing baths to 0.04 in denser, slower baths. We adopt a tolerance of $\epsilon_\infty=0.2$, which here also corresponds to a deviation of about $20\%$ in $T_2$, marked by the line in \ref{fig:validitymap}(c) and the contour in \ref{fig:validitymap}(f). The error grows with the ratio $J_{\mathrm{typ}}/\gamma_{e1}$ of dipolar coupling-to-relaxation. For $J_{\mathrm{typ}}/\gamma_{e1}\lesssim12$ it remains below the tolerance at every sampled depth and below $0.03$ in most sets.  At larger ratios it approaches the tolerance, and at $J_{\mathrm{typ}}/\gamma_{e1}\simeq100$ it crosses it at depths $d\lesssim1.3\,d_{ee}$, by less than twice the sampling error.  For $d\gtrsim1.5\,d_{ee}$ the error stays below the tolerance at every sampled ratio. The error is largest for $d\lesssim d_{ee}$, where the NV is most sensitive to exchange among a few nearby bath spins, and decreases for $d\gg d_{ee}$, where many weakly coupled spins contribute, reducing the sensitivity to individual pair dynamics.

{The ratio $J_{\mathrm{typ}}/\gamma_{e1}$ does not account for the pair detunings or for bath dephasing, which enter the pair parameter $\lambda_{ij}$ through $\Delta_{ij}$ and $\Gamma_{ij}=\gamma_{e1}+2\gamma_{e2}$. Figures~\ref{fig:validitymap}(d,e) isolate these two variables at fixed $J_{\mathrm{typ}}/\gamma_{e1}$. In Fig.~\ref{fig:validitymap}(d), we add to each bath spin a static on-site energy $\varepsilon_i$, drawn from a zero-mean Gaussian of width $\sigma_\varepsilon$, which shifts every pair detuning, $\Delta_{ij}^{(a)}\to\Delta_{ij}^{(a)}+\varepsilon_i-\varepsilon_j$ (Appendix~\ref{meanfieldapproximation}). We characterize the detuning scale by $\Delta_{\mathrm{typ}}$, the geometric mean of $\Delta_{\mathrm{rms},ij}$ over nearest-neighbor pairs, where $\Delta_{\mathrm{rms},ij}^2=\sum_a p_a\langle[\Delta_{ij}^{(a)}(\mathbf s)]^2\rangle_{\mathbf s}$. Without disorder, $\Delta_{\mathrm{typ}}/2\pi\approx0.3$~MHz, set by the probe-induced and spectator fields; as disorder increases it to about $9$~MHz, the deviation decreases by roughly an order of magnitude for all three ratios: a detuned pair exchanges polarization at a rate suppressed as $(J_{ij}/\Delta_{ij}^{(a)})^2$, and its residual dispersive shifts are static on the echo time scale and refocused by the $\pi$ pulse, so the kinetic description becomes more accurate in this detuned regime. In Fig.~\ref{fig:validitymap}(e) the pair linewidth is increased by bath dephasing $\gamma_{e2}$. For all three ratios the deviation first grows and is largest at $\gamma_{e2}/2\pi\simeq20$ to $30$~kHz, where the pair linewidth $\Gamma_{ij}$ becomes comparable to the detunings of the more nearly resonant pairs and their transfer rate, $\overline W_{ij}\propto\Gamma_{ij}/(\Delta_{ij}^2+\Gamma_{ij}^2)$, is largest. It then falls as the pairs become overdamped. In this limit the pair coherence relaxes faster than the populations and the rate reduction becomes a true Markovian description (Appendix~\ref{sec:pair_projection_sm}). Bath spin dephasing thus extends the validity of KCE only once $\gamma_{e2}$ exceeds the pair detunings, whereas bath detuning improves it at every strength.}

We next examine how explicitly propagated spin clusters capture coherent exchange, how exchange shapes the echo decay, and which cluster sizes dominate decoherence. In the singleton-local limit, $\lambda_c\to\infty$, every bath spin forms a singleton, and exchange is represented solely by effective local switching rates; ME-CCE1, which neglects flip-flops, separates the role of flip-flops from that of relaxation. Figure~\ref{fig:purelocal}(a,b,d,e) follows $T_2$ and the stretching exponent $n$, defined by $\mathcal L(t)=\exp[-(t/T_2)^n]$, through depth sweeps at $T_{1e}=100~\mu$s. To determine whether decoherence is dominated by individual spins, coherent pairs, or larger correlated groups as bath density increases, we group the factorized local coherence by cluster size; $G_k$ is the product of the coherence factors of all $k$-spin clusters [Fig.~\ref{fig:purelocal}(c,f), averages over 100 bath realizations at $d=10$~nm]. 

At $\rho=0.001~\mathrm{nm}^{-2}$ ($J_{\mathrm{typ}}/\gamma_{e1}\simeq1.3$), the combined probe coupling $A_{\mathrm{eff}}=(\sum_iA_i^2)^{1/2}$ gives $A_{\mathrm{eff}}/\gamma_{e1}\simeq2.4$--$3.1$ for $d\le20$~nm and $n$ lies between $2.2$ and $2.4$, falling to $1.4$ at $50$~nm as $A_{\mathrm{eff}}/\gamma_{e1}$ drops to $0.55$ [Fig.~\ref{fig:purelocal}(a,b)]. ME-CCE1 lies within $5\%$ of ME-CCE4 in $T_2$, so flip-flops play a minor role here, and network KCE follows ME-CCE4 to within $5\%$ in $T_2$ and $0.05$ in $n$ at every depth. The singleton-local limit nevertheless overestimates $T_2$ at depth, giving about $3.0$ instead of $1.7$~ms at $d=50$~nm. It adds the flip-flop transfer rates $\overline{W}_{ij}$ to the switching rate of every spin; where $A_{\mathrm{eff}}\lesssim\gamma_{e1}$, faster switching narrows the probe-field fluctuations and lengthens $T_2$, whereas a flip-flop of two nearly equally coupled spins hardly changes the field. Replacing exchange by independent switching rates therefore fails at depth, where neglecting it is accurate. In this bath the explicit clusters are almost all singletons and pairs: at $d=10$~nm the singletons carry about $60\%$ of the decay and the rest is by the pairs, while larger clusters hardly contribute [Fig.~\ref{fig:purelocal}(c)].

At $\rho=0.01~\mathrm{nm}^{-2}$ ($J_{\mathrm{typ}}/\gamma_{e1}\simeq42$), flip-flops dominate at shallow depth: ME-CCE1 gives up to about twice the ME-CCE4 $T_2$ and $n\simeq1.8$ instead of $1.3$, and approaches ME-CCE4 beyond $d\simeq14$~nm [Fig.~\ref{fig:purelocal}(d,e)]. The ME-CCE4 exponent rises from about $1.3$ at shallow depth to $2.7$ near $d=18$~nm, and network KCE reproduces this rise to within $0.25$, with $T_2$ within about $15\%$. The flip-flop couplings are the same at every depth; what changes is the field step $\delta A$ that a flip-flop produces at the probe. Near the probe these steps are large, so flip-flops add to the dephasing, and the decay becomes faster and closer to exponential. Far from the probe, the two spins of a pair couple almost equally to it, flip-flops barely change its field, and the decay approaches that of independent spins. The singleton-local limit fails in this bath: it gives a shorter $T_2$ at shallow depth and a longer one at $20$~nm, and it overestimates $n$ at shallow depth, giving about $2.3$ instead of $1.3$, and underestimates it at $20$~nm, giving $1.8$ instead of $2.6$. At $d=18$~nm it even reproduces $T_2$ while giving $n=1.9$ instead of $2.7$. The exponent is thus a sharper test of the partition than $T_2$. At $d=10$~nm the pair factor carries about half of the decay and the four-spin factor about $40\%$, the singletons a smaller share, while the three-spin factor stays close to unity over the decay window [Fig.~\ref{fig:purelocal}(f)]. Coherent clusters, here mainly pairs and four-spin clusters, are therefore the leading correction to the singleton-local limit in the dense bath.

\begin{figure}[!t]
\centering
\includegraphics[width=\columnwidth]{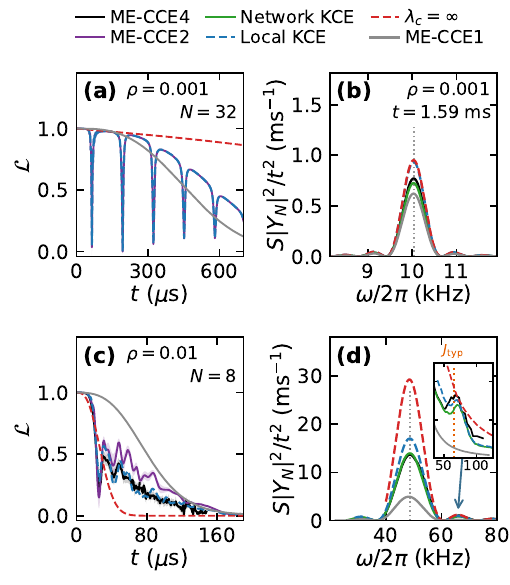}
\caption{Pulse signatures of bath dynamics at $T_{1e}=100~\mu$s for one bath realization. (a,c) CPMG coherence with the NV placed next to a strongly coupled cluster: (a) a pair at $\rho=0.001~\mathrm{nm}^{-2}$, $d=10$~nm, $N=32$, and (c) a four-spin cluster at $\rho=0.01~\mathrm{nm}^{-2}$, $d=10$~nm, $N=8$. Curves show the reference ME-CCE2 and ME-CCE4 (shading: $\pm1$ standard error of the bath-state average) together with local KCE ($\lambda_c=0.5$), the singleton-local limit ($\lambda_c\to\infty$), and ME-CCE1. (b,d) Noise weight $S(\omega)|Y_N(\omega)|^2/t^2$ sampled by the CPMG filter at the decay time $t=T_2$ of ME-CCE4, with the NV at $d=10$~nm: (b) $\rho=0.001~\mathrm{nm}^{-2}$, $N=32$, $t=1.59$~ms, and (d) $\rho=0.01~\mathrm{nm}^{-2}$, $N=8$, $t=82~\mu$s. The spectrum $S(\omega)$ of each method is recovered from its CPMG decays, and network KCE is shown in addition. Grey dotted lines: filter centre $\omega_N=\pi N/t$. Inset of (d): $S(\omega)$ from the decays, with $J_{\mathrm{typ}}$ marked in orange; the arrow marks the second filter lobe, which samples this band.
}

\label{fig:consequence}
\end{figure}

\subsection{Pulse signatures of bath dynamics}
\label{pulses_sig}
The cluster-size decomposition identifies which coherent clusters contribute to dephasing. A pulse sequence can resolve their dynamics. A strongly coupled pair exchanges polarization coherently at a frequency $f$ set mainly by its dipolar coupling, and each flip-flop changes the field at the probe by $\delta A=A_i-A_j$. Under a CPMG sequence of $N$ pulses separated by $2\tau=t/N$, the pulses rectify this field when $\tau=(2k-1)/(4f)$, so the coherence dips at $t=2N\tau$ instead of decaying smoothly. Figure~\ref{fig:consequence}(a) shows this signature for a pair with dipolar coupling $124$~kHz close to an NV in one realization at $\rho=0.001~\mathrm{nm}^{-2}$. Local KCE keeps the pair as an explicit cluster and reproduces the dips of ME-CCE2 and ME-CCE4. ME-CCE1, which neglects flip-flops, shows no dips and decays faster, since it misses both the resonances and the averaging of $\delta A$ by the fast flip-flops. The singleton-local limit, which replaces the pair by an effective rate, also shows no dips, and its decay is far too slow. Figure~\ref{fig:consequence}(c) shows a four-spin cluster at $\rho=0.01~\mathrm{nm}^{-2}$ with internal couplings of $40$ to $70$~kHz, which local KCE keeps as one cluster. Its several flip-flop frequencies produce oscillations rather than isolated dips. Local KCE follows ME-CCE4 to within about its sampling error, whereas ME-CCE2 overestimates the oscillations and the singleton-local limit and ME-CCE1 miss them.

Figure~\ref{fig:consequence}(b,d) shows which part of the bath noise a CPMG sequence samples at the decay time. The filter $|Y_N(\omega)|^2$ weights the noise spectrum $S(\omega)$ near $\omega_N=\pi N/t$, and here $S(\omega)$ of each method is recovered from its CPMG decays (Appendix~\ref{app:cpmgspectra}). In the dilute bath at $N=32$, the filter samples the noise near $10$~kHz, far below the pair frequency, and all methods give this weight to within about $25\%$ [Fig.~\ref{fig:consequence}(b)]. In the dense bath at $N=8$, network KCE gives the weight of ME-CCE4 to within a few percent and local KCE to within about $20\%$, whereas the singleton-local limit gives about twice and ME-CCE1 about a third of it [Fig.~\ref{fig:consequence}(d)]. The second filter lobe falls near $J_{\mathrm{typ}}$, where ME-CCE4 and both KCE formulations show a local maximum of $S(\omega)$. Neither ME-CCE1 nor the singleton-local limit can produce such a maximum: for independently relaxing spins $S(\omega)$ is a sum of zero-frequency Lorentzians and decreases monotonically. These signatures separate coherent exchange from relaxation-driven switching, and reproducing them requires propagating the coherent clusters explicitly, as KCE does.

\begin{figure}[!t]
\centering
\includegraphics[width=0.8\columnwidth]{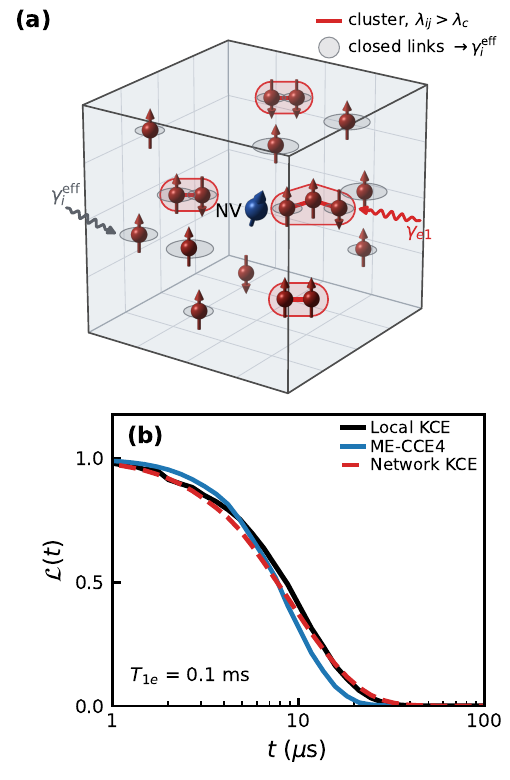}
\caption{
Three-dimensional P1 electron-spin bath at a concentration of $3$~ppm in a $600$~G field along the NV axis.
(a)~Schematic of the local KCE partition (illustrative positions, not to scale): P1 centres throughout the crystal around the NV (blue). Spins joined by $\lambda_{ij}>\lambda_c$ (red) form explicit clusters of at most four spins (outlined) that evolve independently; every other link is closed into the rate $\gamma_i^{\rm eff}$ of both partners (gray discs, size growing with the number of closed links). Red pulse: intrinsic relaxation $\gamma_{e1}$; gray pulse: rates $\gamma_i^{\rm eff}$ from closed links. In one realization (185 bath spins within 45~nm of the NV, $\lambda_c=0.5$), 125 spins form 53 clusters (41 pairs, 5 triples, 7 quadruples) and 60 remain singletons.
(b)~Ensemble-averaged NV Hahn-echo coherence at $T_{1e}=0.1$~ms from local KCE, network KCE, and ME-CCE4.}
\label{fig:p1}

\label{fig:P1}
\end{figure}

\subsection{Extension to a three-dimensional P1 bath}
\label{P1_cent}
Substitutional nitrogen (P1) centers inside diamond provide a
three-dimensional test of KCE. Their nuclear hyperfine interactions
shift the electron-spin energies and affect which pairs can exchange
polarization~\cite{park2022decoherence}. We include this shift as a static on-site energy $\varepsilon_i$ of each P1 electron spin (Appendix~\ref{meanfieldapproximation}), using the $^{14}$N hyperfine tensors of Ref.~\cite{park2022decoherence} for a Jahn--Teller axis assigned at random among the four $\langle111\rangle$ orientations and a randomly sampled nuclear-spin projection $m_I\in\{0,\pm1\}$. Previous calculations
have used CCE4 for NV echo~\cite{marcks2024guiding}; it has been shown that correlations
among larger spin groups become important at higher P1 densities
and under longer CPMG sequences~\cite{park2026quantum}. Including
bath dissipation makes propagation of these higher-order clusters
especially costly. 
Reported room-temperature single-P1 polarization-decay times include $(75\pm20)~\mu\mathrm{s}$ in nitrogen-rich type-Ib diamond at $B=600~\mathrm{G}$~\cite{hanson2006polarization} and $(1.9\pm0.2)~\mathrm{ms}$ in Ref.~\cite{marcks2024quantum}. Local polarization decay can include spin--spin contributions and therefore need not directly measure intrinsic spin--lattice relaxation~\cite{hanson2006polarization}. For the dissipative benchmark, we adopt $T_{1e}=0.1~\mathrm{ms}$ as a phenomenological local longitudinal-relaxation time. We compare KCE and ME-CCE4 using the same bath and dissipative model to assess whether the clusters selected by $\lambda_{ij}$ reproduce the NV echo. We compare KCE with ME-CCE4 for the same bath and dissipative model to access whether the clusters selected by $\lambda_{ij}$ suffice to reproduce the NV echo.

Figure~\ref{fig:p1}(a) shows how the pair criterion partitions
one bath realization into multispin clusters and singletons.
For the ensemble-averaged echo in Fig.~\ref{fig:p1}(b), the local KCE at $\lambda_{c}=0.5$
gives $T_2=10.8~\mu\mathrm{s}$, compared with
$9.3~\mu\mathrm{s}$ from ME-CCE4. Network KCE, which retains
joint transfers between the clusters, gives
$T_2=10.1~\mu\mathrm{s}$. Thus, local KCE reproduces the
ME-CCE4 coherence time to within $16\%$ in this P1 bath; retaining the kinetic network reduces the difference
to $9\%$.

\section{Discussion}

We have developed kinetic cluster embedding (KCE), a hybrid quantum--classical description of qubit decoherence in a dissipative, interacting spin bath. A single pair parameter $\lambda_{ij}$, which compares the flip-flop coupling of two bath spins with their detuning and linewidth, divides the bath into near-resonant, strongly coupled groups, propagated with the full dissipative quantum dynamics; isolated relaxing spins, treated through their exact equivalence to random-telegraph fields \cite{bergli2007exact, breuer2002theory}; and weakly coupled pairs, represented by polarization-transfer rates in a kinetic network.

Our central result is that coupling strength alone does not determine how bath spins affect the probe. What matters is whether a pair can complete an appreciable part of an exchange oscillation before detuning and damping suppress it. If it cannot, polarization is transferred incoherently, as random jumps; if it can, the exchange is coherent and must be treated quantum mechanically. Detuning suppresses exchange and improves the rate description at every strength, whereas dephasing degrades it until the pair becomes overdamped. For the same reason, the accuracy of KCE does not simply decrease as $J_{\mathrm{typ}}/\gamma_{e1}$ grows: network KCE follows ME-CCE4 from relaxation-dominated baths to baths in which the dipolar coupling exceeds relaxation up to $\sim10^{2}$, with $T_2$ typically within $15\%$, and larger deviations only for shallow NVs, $d\leq d_{ee}$ i.e., when the NV is closer to the surface than the mean spacing between surface spins. The partition also shows how exchange affects the probe. In dense baths, coherent clusters are the leading correction to independent fluctuators with flip-flop-renormalized rates. Their clearest signature is the response to pulses: under CPMG control, a coherent pair produces dips at its flip-flop frequency, and a larger cluster produces oscillations. The echo exponent is less specific, since relaxation alone already makes the decay non-exponential, but it remains a sharper test of the partition than $T_2$ (Sec.~\ref{bench}).

Beyond the clusters, the kinetic network retains a correlation that local pictures omit: polarization transferred between clusters is conserved and can return before it relaxes. At fast bath relaxation this return is erased and the local closure suffices. At slow relaxation it controls the coherence: with the same cluster partition, the local closure, which discards it, deviates from ME-CCE4 by more than 25\% of $T_2$, whereas network KCE follows the reference (Appendix~\ref{benchmarks}). Polarization return is thus a distinct source of bath correlations, and on larger scales the kinetic network could be coarse-grained into a spin-diffusion description of collective polarization transport.

Computationally, KCE propagates only the spins that require it: in the densest surface bath studied, at most about seven four-spin clusters per realization instead of more than $10^5$ in ME-CCE4, with the local closure running more than three orders of magnitude faster (Appendix~\ref{sec:clusterCountsSM}). The disjoint partition also avoids an instability of cluster-correlation expansions. ME-CCE evaluates the coherence as $\mathcal L=\prod_C\mathcal L_C^{\,p_C}$ over all clusters up to the truncation order, with $p_C=1-\sum_{C'\supset C}p_{C'}$, so some cluster contained in larger retained clusters enters with a negative power. For a sampled pure bath state, the coherence of a strongly coupled subcluster can pass close to zero at a dip of its exchange oscillation; the larger clusters that would cancel this pole in the full expansion are only partly retained, and the truncated product can grow without bound, as found for sparse dipolar baths \cite{witzel2012quantum, schaetzle2024spin}.  In KCE every cluster factor enters with power $+1$ [Eq.~\eqref{eq:Lhybrid_main}], so no subcluster is divided out.

KCE reaches its own limit when the cutoff needed for a given accuracy makes the strong-link graph percolate, as expected for dense, weakly disordered, or slowly relaxing baths such as nearly ordered spin lattices. A percolating cluster signals extended coherent dynamics that requires higher-order ME-CCE or a full many-body treatment. By organizing bath dynamics through the local competition between exchange, detuning, and damping, KCE provides a framework for exploring decoherence in other dissipative spin environments, including other color-center, donor, and molecular spin baths \cite{Seth2026, Diego2025, ryan2026challenge} and for using local probes to distinguish coherent exchange from incoherent polarization transport in disordered dipolar quantum simulators \cite{yan2013observation}.

\vspace*{0.5em}
\textit{Codes used.} The conventional CCE and the master-equation CCE (ME-CCE) are available in the PyCCE package \cite{onizhuk2021pycce}; the kinetic cluster embedding (KCE) is implemented therein. 

\vspace*{0.5em}
\textit{Acknowledgments}
This work was supported by the NSF QuBBE Quantum Leap Challenge Institute (Grant No. NSF OMA-2121044)
and was carried out with the use of resources provided by the University of Chicago's Research Computing Center (RCC).

\appendix
\setcounter{secnumdepth}{3}

\section{\label{Noneequalrate}General single-bath-spin solution: unequal rates and arbitrary initial bath state}

We consider the central spin $v$ longitudinally coupled to a single bath spin $e$ with coupling strength $A$. In the toggling frame of the chosen central-spin control sequence, the Hamiltonian is given by: 

\begin{equation} 
\hat H_N(t)=\frac{A}{4}\,y_N(t)\,\hat\sigma_z^{(v)}\hat\sigma_z^{(e)}, \label{eq:toggledhamiltonian} 
\end{equation}

where $y_N(t)=\pm1$ is the toggling function of a sequence of $N$ ideal central-spin $\pi$ pulses, which changes sign at each pulse.

In the presence of dissipation, $\gamma_{v1}$ and $\gamma_{v2}$ denote the longitudinal-relaxation and pure-dephasing rates of the central spin, respectively, while $\gamma_{e1}$ and $\gamma_{e2}$ denote the corresponding rates of the bath spin. To allow for a polarized bath, we resolve the bath longitudinal-relaxation channel into the physical transition rates

\begin{equation*}
  \gamma_{e\uparrow}:\ |\downarrow_e\rangle\rightarrow|\uparrow_e\rangle,
  \qquad
  \gamma_{e\downarrow}:\ |\uparrow_e\rangle\rightarrow|\downarrow_e\rangle,
\end{equation*}

so that 
\begin{equation*}
  \gamma_{e1}\equiv\gamma_{e\uparrow}+\gamma_{e\downarrow},
  \qquad
  \Delta\gamma_e\equiv\gamma_{e\uparrow}-\gamma_{e\downarrow}.
\end{equation*}

The symmetric-switching model of the main text is recovered when $\gamma_{e\uparrow}=\gamma_{e\downarrow}=\gamma_{e1}/2$, for which $\Delta\gamma_e=0$.

The joint density operator $\hat\rho_N(t)$ therefore evolves as:  
\begin{equation}
\begin{aligned}
  \dot{\hat\rho}_N={}&-i[\hat H_N(t),\hat\rho_N]
  \\
  &\quad
  +\frac{\gamma_{v1}}{2}\mathcal D[\hat\sigma_+^{(v)}]\hat\rho_N
  +\frac{\gamma_{v1}}{2}\mathcal D[\hat\sigma_-^{(v)}]\hat\rho_N
  \\
  &\quad
  +\frac{\gamma_{v2}}{2}\mathcal D[\hat\sigma_z^{(v)}]\hat\rho_N
  \\
  &\quad+\gamma_{e\uparrow}\mathcal D[\hat\sigma_+^{(e)}]\hat\rho_N
  +\gamma_{e\downarrow}\mathcal D[\hat\sigma_-^{(e)}]\hat\rho_N
  \\
  &\quad
  +\frac{\gamma_{e2}}{2}\mathcal D[\hat\sigma_z^{(e)}]\hat\rho_N.
\end{aligned}
  \label{eq:appendix_master_general}
\end{equation}

where, $\mathcal D[\hat L]\hat\rho
  =\hat L\hat\rho\hat L^\dagger
  -\frac{1}{2}\left\{\hat L^\dagger\hat L,\hat\rho\right\}$
is the Lindblad dissipator.

We take the initial joint state to be factorized, with the central spin $v$ prepared along $+x$,
\begin{equation}
  \hat\rho_N(0)=\hat\rho_v(0)\otimes\hat\rho_e(0),
  \qquad
  \hat\rho_v(0)
  =\frac{1}{2}\left(\hat\sigma_0^{(v)}+\hat\sigma_x^{(v)}\right).
\end{equation}
The initial bath density matrix is otherwise arbitrary and can be written in Bloch form \cite{nielsen2000quantum} as
\begin{equation}
\begin{gathered}
  \hat\rho_e(0)
  =\frac{1}{2}\left(
    \hat\sigma_0^{(e)}
    +\alpha\hat\sigma_z^{(e)}
    +\beta\hat\sigma_x^{(e)}
    +\eta\hat\sigma_y^{(e)}
  \right),
  \\
  \alpha^2+\beta^2+\eta^2\leq1.
\end{gathered}
\end{equation}
The parameters $\alpha$, $\beta$, and $\eta$ are the initial $z$, $x$, and $y$ components of the bath Bloch vector, respectively.

Thus the central-spin coherence is contained in the off-diagonal block
$ \hat\tau_N(t)\equiv
 \langle0_v|\hat\rho_N(t)|1_v\rangle,
$ which is an operator on the bath-spin Hilbert space.  Expanding this operator in the bath Pauli basis gives

\begin{equation}
\hat\tau_N(t)=\frac14\left(
 u_N\hat\sigma_0^{(e)}+m_N\hat\sigma_z^{(e)}
+p_N\hat\sigma_x^{(e)}+q_N\hat\sigma_y^{(e)}\right),
\label{eq:appendix_tau_general}
\end{equation}

where the four generally complex coefficients are defined by
\begin{align*}
  u_N&=2\operatorname{Tr}_e[\hat\tau_N],
  &m_N&=2\operatorname{Tr}_e[\hat\sigma_z^{(e)}\hat\tau_N],\\
  p_N&=2\operatorname{Tr}_e[\hat\sigma_x^{(e)}\hat\tau_N],
  &q_N&=2\operatorname{Tr}_e[\hat\sigma_y^{(e)}\hat\tau_N].
\end{align*}

In the bath basis $\{|\uparrow_e\rangle,|\downarrow_e\rangle\}$,
\begin{equation}
  \hat\tau_N
  =\frac{1}{4}
  \begin{pmatrix}
    u_N+m_N & p_N-iq_N\\
    p_N+iq_N & u_N-m_N
  \end{pmatrix}.
\end{equation}

For the factorized initial state specified above, 
  $\hat\tau_N(0)
  =\langle0_v|\hat\rho_v(0)|1_v\rangle\,\hat\rho_e(0)
  =\hat\rho_e(0)/2$.  Equation~\eqref{eq:appendix_tau_general} therefore gives
  $u_N(0)=1,
 m_N(0)=\alpha,
 p_N(0)=\beta,
 q_N(0)=\eta.$
Moreover, Eq.~\eqref{eq:appendix_tau_general} gives $\operatorname{Tr}_e\hat\tau_N(t)=u_N(t)/2$ and $\operatorname{Tr}_e\hat\tau_N(0)=1/2$.  The coherence function defined in the main text is therefore
\begin{equation}
  \mathcal L_N(t)
  \equiv
  \operatorname{Re}\!\left[
    \frac{\operatorname{Tr}_e\hat\tau_N(t)}
         {\operatorname{Tr}_e\hat\tau_N(0)}
  \right]
  =\operatorname{Re}u_N(t).
\end{equation}

To obtain the equations of motion, note first that the Hamiltonian contribution to the off-diagonal block is
\begin{equation*}
  \dot{\hat\tau}_N\big|_H
  =-\frac{iA}{4}y_N(t)
  \bigl\{\hat\sigma_z^{(e)},\hat\tau_N\bigr\}.
\end{equation*}

Combining this with the dissipative terms in Eq.~\eqref{eq:appendix_master_general}, gives

\begin{align}
  \dot u_N
  &=-\Gamma_vu_N-\frac{iA}{2}y_Nm_N,
  \label{eq:appendix_u_eq}\\
  \dot m_N
  &=-\left(\Gamma_v+\gamma_{e1}\right)m_N
    +\Delta\gamma_e\,u_N
    -\frac{iA}{2}y_Nu_N,
  \label{eq:appendix_m_eq}\\
  \dot p_N
  &=-\left(\Gamma_v+\frac{\gamma_{e1}}{2}+\gamma_{e2}\right)p_N,
 \label{eq:appendix_p_eq}\\
  \dot q_N
  &=-\left(\Gamma_v+\frac{\gamma_{e1}}{2}+\gamma_{e2}\right)q_N,
  \label{eq:appendix_q_eq}
\end{align}

where $\Gamma_v\equiv\gamma_{v1}/2+\gamma_{v2}$ is the intrinsic central-spin coherence-decay rate. Equations~\eqref{eq:appendix_u_eq} and \eqref{eq:appendix_m_eq} form a closed sector: their right-hand sides depend only on $u_N$ and $m_N$.  The transverse amplitudes $p_N$ and $q_N$ do not enter $u_N$ and therefore do not affect $\mathcal L_N(t)$ for the longitudinal coupling considered here.  Consequently, the coherence function is independent of the initial transverse bath components $\beta$ and $\eta$, as well as the bath pure-dephasing rate $\gamma_{e2}$.

Therefore, to factor out the intrinsic central-spin coherence decay, we define:
\begin{equation}
  u_N(t)=e^{-\Gamma_v t}\widetilde u_N(t),
  \qquad
  m_N(t)=e^{-\Gamma_v t}\widetilde m_N(t).
\end{equation}
The resulting amplitudes satisfy

\begin{equation}
\begin{gathered}
  \frac{d}{dt}
  \begin{pmatrix}
    \widetilde u_N\\
    \widetilde m_N
  \end{pmatrix}
  =M[y_N(t)]
  \begin{pmatrix}
    \widetilde u_N\\
    \widetilde m_N
  \end{pmatrix},
  \\
  M[y]\equiv
  \begin{pmatrix}
    0&-iAy/2\\
    \Delta\gamma_e-iAy/2&-\gamma_{e1}
  \end{pmatrix}.
\end{gathered}
  \label{eq:appendix_matrix_general}
\end{equation}

For symmetric bath relaxation, $\Delta\gamma_e=0$, Eq.~\eqref{eq:appendix_matrix_general} reduces directly to Eq.~\eqref{eq:reduced_2x2_system} of the main text.

To make the connection between Eq.~\eqref{eq:appendix_matrix_general} and a classical two-state fluctuator, we examine the diagonal matrix elements of $\hat\tau_N(t)$ in the bath $\hat\sigma_z^{(e)}$ basis, given as:

\begin{align}
  c_+(t)
  &\equiv 2e^{\Gamma_v t}
  \langle\uparrow_e|\hat\tau_N(t)|\uparrow_e\rangle
  =\frac{\widetilde u_N(t)+\widetilde m_N(t)}{2},\\
  c_-(t)
  &\equiv 2e^{\Gamma_v t}
  \langle\downarrow_e|\hat\tau_N(t)|\downarrow_e\rangle
  =\frac{\widetilde u_N(t)-\widetilde m_N(t)}{2}.
\end{align}

Their sum therefore gives the central-spin coherence, $
\widetilde u_N(t)=c_+(t)+c_-(t)$. For the product initial state, $\hat\tau_N(0)=\hat\rho_e(0)/2$, and their initial values are
$
  c_+(0)=\langle\uparrow_e|\hat\rho_e(0)|\uparrow_e\rangle
  =\frac{1+\alpha}{2} \ \text{and} \
  c_-(0)=\langle\downarrow_e|\hat\rho_e(0)|\downarrow_e\rangle
  =\frac{1-\alpha}{2}.
$
where $\alpha=\operatorname{Tr}_e[\hat\sigma_z^{(e)}\hat\rho_e(0)]$ is the initial longitudinal polarization of the bath spin. Thus, only at $t=0$ do $c_+(t)$ and $c_-(t)$ coincide with the populations of $|\uparrow_e\rangle$ and $|\downarrow_e\rangle$, respectively. Substituting $c_+(t)$ and $c_-(t)$ into Eq.~\eqref{eq:appendix_matrix_general} gives
\begin{align}
  \dot c_+
  &=-\frac{iA}{2}y_N(t)c_+
    -\gamma_{e\downarrow}c_+
    +\gamma_{e\uparrow}c_-,
  \label{eq:up_eq}\\
  \dot c_-
  &=+\frac{iA}{2}y_N(t)c_-
    +\gamma_{e\downarrow}c_+
    -\gamma_{e\uparrow}c_-.
  \label{eq:down_eq}
\end{align}

Equations~\eqref{eq:up_eq} and ~\eqref{eq:down_eq}  have the same form as the probability-weighted conditional-coherence equations for a classical random-telegraph field $\xi(t)=\pm A/2$ \cite{breuer2002theory}. In this classical representation, $c_+(t)$ and $c_-(t)$ are the ensemble-averaged central-spin coherence contributions conditioned on the fluctuator having the values $+A/2$ and $-A/2$, respectively, at time $t$. The transitions $+\rightarrow-$ and $-\rightarrow+$ occur at rates $\gamma_{e\downarrow}$ and $\gamma_{e\uparrow}$, respectively. Thus, for the central-spin coherence, an arbitrary initial bath state is equivalent to a classical initial distribution determined only by $\alpha$. This mapping concerns the central-spin coherence and does not imply an equivalence between the complete quantum bath state and a classical probability distribution.

We now use Eq.~\eqref{eq:appendix_matrix_general}
to obtain the central-spin coherence for an arbitrary pulse
sequence. The evolution is governed by the time-ordered propagator

\begin{equation}
  \mathcal U_N(t)
  \equiv\mathcal T\exp\!\left(
    \int_0^t M[y_N(s)]\,ds
  \right),
\end{equation}

For compactness, we write the initial conditions
$\widetilde u_N(0)=1$ and $\widetilde m_N(0)=\alpha$ as $
  \mathbf v_0
  \equiv
  \begin{pmatrix}
    1\\
    \alpha
  \end{pmatrix},$ evolving as
 $
  \begin{pmatrix}
    \widetilde u_N(t)\\
    \widetilde m_N(t)
  \end{pmatrix}
  =\mathcal U_N(t)\mathbf v_0.
$ Since the central-spin coherence is determined by   $
  \widetilde u_N(t)
  =\mathbf r^{\mathsf T}\mathcal U_N(t)\mathbf v_0 
$ where
$\mathbf r^{\mathsf T}\equiv(1,0)$, combining this result with $\mathcal L_N(t)=e^{-\Gamma_v t}\operatorname{Re}\widetilde u_N(t)$ gives the exact coherence
\begin{equation}
  \mathcal L_N(t)
  =e^{-\Gamma_v t}\operatorname{Re}\!\left[
    \mathbf r^{\mathsf T}\mathcal U_N(t)\mathbf v_0
  \right].
\end{equation}

For Ramsey evolution, $N=0$ and $y_0(t)=+1$, so the matrix
$M[+1]$ in Eq.~\eqref{eq:appendix_matrix_general} is time independent with eigenvalues
$-\gamma_{e1}/2\pm\Omega_e/2$,  where $ \Omega_e \equiv\sqrt{\gamma_{e1}^2-A^2-2iA\Delta\gamma_e}$.

The corresponding solution for $\widetilde u_0(t)$ is
\begin{equation*}
  \widetilde u_0(t)
  =e^{-\gamma_{e1}t/2}
  \left[
    \cosh\!\left(\frac{\Omega_e t}{2}\right)
    +\frac{\gamma_{e1}-iA\alpha}{\Omega_e}
    \sinh\!\left(\frac{\Omega_e t}{2}\right)
  \right]
\end{equation*}
and hence
\begin{equation}
\begin{aligned}
  \widetilde{\mathcal L}_0(t)
  ={}&\operatorname{Re}\!\Bigl\{
    e^{-\gamma_{e1}t/2}\\
    &\times\Bigl[
      \cosh\!\left(\frac{\Omega_e t}{2}\right)
      +\frac{\gamma_{e1}-iA\alpha}{\Omega_e}
      \sinh\!\left(\frac{\Omega_e t}{2}\right)
    \Bigr]
  \Bigr\}.
\end{aligned}
  \label{eq:rammsey}
\end{equation}
The full Ramsey coherence is $\mathcal L_0(t)=e^{-\Gamma_v t}\widetilde{\mathcal L}_0(t)$.

For the symmetric bath-relaxation model  $
  \Delta\gamma_e=0$ as in the main text, with the bath initially unpolarized, $\alpha=0$, Eq.~\eqref{eq:rammsey} becomes
\begin{equation}
\begin{gathered}
  \widetilde{\mathcal L}_0(t)
  =e^{-\gamma_{e1}t/2}
  \left[
    \cosh\!\left(\frac{\kappa t}{2}\right)
    +\frac{\gamma_{e1}}{\kappa}
    \sinh\!\left(\frac{\kappa t}{2}\right)
  \right],
  \\
  \kappa\equiv\sqrt{\gamma_{e1}^2-A^2}.
\end{gathered}
  \label{eq:ramsey_symmetric}
\end{equation}

For a standard $N$-pulse CPMG sequence, let
\begin{equation*}
\begin{aligned}
  \delta&\equiv\frac{t}{2N},
  \qquad
  s_k\equiv(-1)^k,
  \\
  \Delta_0=\Delta_N&=\delta,
  \\
  \Delta_k&=2\delta\quad(1\leq k\leq N-1).
\end{aligned}
\end{equation*}

The pulse times are $t_j=(2j-1)\delta$, so the first and last free-evolution intervals have duration $\delta$, while all internal intervals have duration $2\delta$.  Since the toggling function is constant on each interval, Eq.~\eqref{eq:appendix_matrix_general} becomes

\begin{equation}
\begin{gathered}
  \mathcal L_N(t)
  =e^{-\Gamma_vt}\operatorname{Re}\!\left[
    \mathbf r^{\mathsf T}\mathcal U_N(t)\mathbf v_0
  \right],
  \\
  \mathcal U_N(t)
  =e^{M[s_N]\Delta_N}\cdots e^{M[s_1]\Delta_1}e^{M[s_0]\Delta_0}.
\end{gathered}
\end{equation}

For symmetric relaxation and an initially unpolarized bath spin,
\begin{equation}
\begin{aligned}
  \widetilde{\mathcal L}_N(t)
  ={}&\frac{e^{-\gamma_{e1}t/2}}{2R_N}
  \\
  &\times\Bigl[
    (R_N+b_N)(a_N+R_N)^N
  \\
  &\qquad+(R_N-b_N)(a_N-R_N)^N
  \Bigr],
\end{aligned}
\label{eq:cpmg_closed}
\end{equation}
where $\omega=\sqrt{A^2-\gamma_{e1}^2}$,
$a_N=(\gamma_{e1}/\omega)\sin(\omega\delta)$,
$b_N=1+(2\gamma_{e1}^2/\omega^2)\sin^2(\omega\delta/2)$,
and $R_N=\sqrt{1+a_N^2}$.
The full CPMG coherence is
$\mathcal L_N(t)=e^{-\Gamma_vt}\widetilde{\mathcal L}_N(t)$. For $|A|<\gamma_{e1}$, Eq.~\eqref{eq:cpmg_closed} is understood by analytic continuation, $\omega=i\sqrt{\gamma_{e1}^2-A^2}$.

For Hahn echo, $N=1$, and this reduces to
\begin{equation}
\begin{aligned}
  \widetilde{\mathcal L}_1(t)
  ={}&e^{-\gamma_{e1}t/2}
  \Bigl[
    1
    +\frac{\gamma_{e1}}{\omega}
      \sin\!\left(\frac{\omega t}{2}\right)
    \\
    &\quad+\frac{2\gamma_{e1}^2}{\omega^2}
      \sin^2\!\left(\frac{\omega t}{4}\right)
  \Bigr].
\end{aligned}
  \label{eq:hahnn}
\end{equation}

Equation~\eqref{eq:hahnn}  also makes clear why a relaxing bath spin is not perfectly refocused.  A Hahn-echo pulse reverses the sign of the central spin's coupling through $y_1(t)$, but a stochastic bath transition reverses the bath field independently of the pulse.  A jump during the sequence therefore produces a nonzero accumulated phase even though the static contribution is refocused.

\section{When the bath autocorrelation is sufficient}
\label{connectiontospectrum}

The exact two-state process derived above has the autocorrelation
\begin{equation}
\langle\xi(t)\xi(0)\rangle=\Delta^2e^{-\gamma_c|t|},
\qquad
\Delta^2=\frac{A^2}{4},\quad \gamma_c=\gamma_{e1},
\label{eq:lorentzian_corr}
\end{equation}
for symmetric switching in the stationary state. Its Fourier transform is
\begin{equation}
\widetilde S(\omega)=\frac{\Delta^2}{\pi}\frac{\gamma_c}{\gamma_c^2+\omega^2}.
\label{eq:lorentzian_spectrum}
\end{equation}
 The Lorentzian form of Eq.~\eqref{eq:lorentzian_corr} characterizes only the two-point correlation \cite{Young2012}. The underlying telegraph process is non-Gaussian, with higher-order correlations not fixed by Eq.~\eqref{eq:lorentzian_corr}.

If the accumulated phase is well described by its second cumulant, the coherence takes the spectrum-only form
\begin{equation}
\mathcal L(t)=e^{-\chi(t)},\qquad
\chi(t)=\int_{-\infty}^{\infty}d\omega\,\widetilde S(\omega)\mathcal F(t,\omega),
\label{eq:chi_general}
\end{equation}
where $\mathcal F$ is the filter function of the control sequence \cite{degen2017quantum,biercuk2011dynamical}. For Ramsey and Hahn echo,
\begin{align}
\mathcal F_{\mathrm R}(t,\omega)&=\frac12\frac{\sin^2(\omega t/2)}{(\omega/2)^2},
\label{eq:filter_ramsey}\\
\mathcal F_{\mathrm E}(t,\omega)&=\frac12\frac{\sin^4(\omega t/4)}{(\omega/4)^2},
\label{eq:filter_echo}
\end{align}
which give
\begin{align}
\chi_{\mathrm R}(t)&=\frac{\Delta^2}{\gamma_c^2}\left(\gamma_ct+e^{-\gamma_ct}-1\right),
\label{eq:chi_ramsey}\\
\chi_{\mathrm E}(t)&=\frac{\Delta^2}{\gamma_c^2}\left(\gamma_ct-3+4e^{-\gamma_ct/2}-e^{-\gamma_ct}\right).
\label{eq:chi_echo}
\end{align}

These expressions follow directly from the exact spin solution when a single switch produces only a small phase. Expanding Eqs.~\eqref{eq:ramsey_symmetric} and \eqref{eq:hahnn} for $|A|/\gamma_{e1}\ll1$ gives
\begin{align}
\tilde{\mathcal L}_{0}(t)&=1-\frac{A^2}{4\gamma_{e1}^2}
\left(\gamma_{e1}t+e^{-\gamma_{e1}t}-1\right)
+\mathcal O(A^4/\gamma_{e1}^4),
\label{eq:L0_fastbath}\\
\tilde{\mathcal L}_{1}(t)&=1-\frac{A^2}{4\gamma_{e1}^2}
\Big(\gamma_{e1}t-3+4e^{-\gamma_{e1}t/2}
\notag\\[-2pt]
&\hspace{5.9em}{}-e^{-\gamma_{e1}t}\Big)
+\mathcal O(A^4/\gamma_{e1}^4).
\label{eq:L1_fastbath}
\end{align}
They are the leading expansions of $e^{-\chi_{\mathrm R}}$ and $e^{-\chi_{\mathrm E}}$ with the identification in Eq.~\eqref{eq:lorentzian_corr}. Thus the exact telegraph process and the spectrum-only calculation have the same two-point correlation at every $A/\gamma_{e1}$, but they give the same coherence only when the higher-order phase correlations are negligible. In the long-time fast-switching limit, both reduce to
\begin{equation}
\tilde{\mathcal L}(t)\simeq\exp\!\left(-\frac{A^2}{4\gamma_{e1}}t\right).
\label{eq:motional_narrowing_rate}
\end{equation}

The slow-switching limit is more transparent in the jump picture. For standard CPMG the integrated switching function $Y(s)=\int_0^s y_N(s')ds'$ is a triangular wave of amplitude $\delta=t/(2N)$. A bath flip at time $s$ changes the accumulated phase by $A Y(s)$. Keeping trajectories with zero or one bath flip gives
\begin{equation}
\tilde{\mathcal L}_{N}(t)=e^{-\gamma_{e1}t/2}
\left[1+\frac{\gamma_{e1}N}{A}\sin\!\left(\frac{At}{2N}\right)\right]
+\mathcal O\!\left[(\gamma_{e1}t)^2\right].
\label{eq:LN_slowbath}
\end{equation}
Equation~\eqref{eq:LN_slowbath} is a low-jump expansion, valid for $\gamma_{e1}t\ll1$; it is not a uniform approximation at the full coherence time. The pulse-control crossover follows more generally from the phase accumulated over one interpulse interval,
\begin{equation}
\eta_N(t)=\frac{|A|t}{2N}.
\label{eq:interpulse_phase}
\end{equation}
For a strongly coupled fluctuator the slow-bath envelope decays on the scale $t\sim2/\gamma_{e1}$. Requiring $\eta_N\lesssim1$ on this timescale gives the order-of-magnitude criterion
\begin{equation}
N\gtrsim\frac{|A|}{\gamma_{e1}}.
\label{eq:dd_threshold}
\end{equation}
This criterion identifies a crossover, not a sharp threshold. It expresses the simple requirement that the control reverse the probe phase before a fixed bath field can generate an order-one phase.

The familiar CPMG scaling follows from a different, short-time limit. Because CPMG has zero net area, the static part of the correlation is refocused. Expanding $e^{-\gamma_{e1}|t_1-t_2|}$ over one interpulse interval,
$\gamma_{e1}t/2N\ll1$, with no restriction on $\gamma_{e1}t$, the second
cumulant gives

\begin{equation}
-\ln\tilde{\mathcal L}_N(t)
=\frac{A^2\gamma_{e1}}{48N^2}t^3
+\mathcal O\!\left(\frac{A^2\gamma_{e1}^2t^4}{N^2}\right),
\label{eq:LN_shorttime_revised}
\end{equation}

where the coefficient follows from $\int_0^tY^2(s)ds=t^3/(12N^2)$. The
second cumulant is accurate when many flips occur during the decay,
$\gamma_{e1}t\gg1$, and each contributes a small phase, $|A|t/2N\ll1$,
so that the accumulated phase is Gaussian. Setting the exponent to one,
\begin{equation}
T_2\simeq\left(\frac{48N^2}{A^2\gamma_{e1}}\right)^{1/3}\propto N^{2/3},
\label{eq:T2_N23_revised}
\end{equation}

and both conditions hold at $T_2$ precisely when $N\gg|A|/\gamma_{e1}$,
the crossover of Eq.~\eqref{eq:dd_threshold}. Below it the coherence
time of a strongly coupled fluctuator stays near $2/\gamma_{e1}$;
above it the exact telegraph solution follows
Eq.~\eqref{eq:T2_N23_revised} to within a few percent. The
$N^{2/3}$ law is therefore a fast-pulsing result that applies beyond
the crossover, not a short-time one.

\section{Noninteracting dissipative bath}\label{sm:noninteracting}
When bath-bath couplings are negligible, the bath reduces to a
collection of independent dissipative fluctuators. Let
\(\hat\tau_N^{(i)}(t)\) be the single-bath-spin coherence operator of
the main text for coupling \(A_i\), and we define \(\hat Q_N^{(i)}(t)=2e^{\Gamma_v t}\hat\tau_N^{(i)}(t)\), the
single-spin coherence operator with the probe's intrinsic decay,
\(\Gamma_v=\gamma_{v1}/2+\gamma_{v2}\), factored out, so that
\(\Tr_i\hat Q_N^{(i)}(t)=\widetilde u_N(t)\) and
\(\hat Q_N^{(i)}(0)=\hat\rho_i(0)\). For an initially
uncorrelated bath the coherence block factorizes as
\begin{equation}
\hat\tau_N(t)
=\frac12 e^{-\Gamma_v t}
 \bigotimes_{i=1}^{n}\hat Q_N^{(i)}(t).
\label{eq:tau_factorized_uniform}
\end{equation}
Taking the bath trace, with \(\widetilde u_N(A_i,t)=\Tr_i\hat Q_N^{(i)}(t)\)
the complex single-spin amplitude of Eq.~\eqref{eq:reduced_2x2_system}
for coupling \(A_i\), yields
\begin{equation}
\mathcal L_N\!\left(t\,\middle|\,\{A_i\}\right)
=
e^{-\Gamma_v t}
\operatorname{Re}\!\left[\prod_{i=1}^{n}\widetilde u_N(A_i,t)\right].
\label{eq:factorized_coherence_uniform}
\end{equation}
For an unpolarized bath with symmetric rates,
\(\gamma_{e\uparrow}=\gamma_{e\downarrow}\), every \(\widetilde u_N\) is
real and the coherence reduces to the product of the real single-spin
responses \(\tilde{\mathcal L}_N(A_i,t)=\operatorname{Re}\widetilde u_N(A_i,t)\)
used below; for a polarized bath or unequal rates the amplitudes carry
phases, and the real part must be taken after the product.
Equation~\eqref{eq:factorized_coherence_uniform} applies both to a
static dilute bath and to a random-hopping bath, provided that the
hopping dynamics can still be represented as independent local
dissipative fluctuators. In the language of the main text, the
noninteracting bath is an ensemble of independent classical telegraph
fluctuators; the factorized coherence is the product of non-Gaussian
single-fluctuator responses, which need not be Gaussian even in
aggregate at shallow depths where a few near spins dominate.

We now specialize to a two-dimensional random bath of areal density \(\rho\), as relevant for shallow NV centers coupled to surface spins. For a bath spin located at in-plane position \(\mathbf R\), the coupling to a central spin at depth \(d\) is denoted by \(A(\mathbf R;d)\). Averaging Eq.~\eqref{eq:factorized_coherence_uniform} over bath configurations drawn from a Poisson point process gives
\begin{equation}
\left\langle  \mathcal L_N(t)\right\rangle
=
e^{-\Gamma_v t}
\operatorname{Re} \Bigg[\exp\!\left[
\rho\int d^2R\,
\bigl(
\widetilde u_N(A(\mathbf R;d),t)-1
\bigr)
\right]\Bigg],
\label{eq:config_average_uniform}
\end{equation}
A derivation of Eq.~\eqref{eq:config_average_uniform} is given in Appendix~\ref{C-average}.

Equation~\eqref{eq:config_average_uniform} connects the exact single-spin solution to an experimentally relevant continuum bath. In general, the depth dependence of the configurationally averaged coherence is determined by the single-spin amplitude \(\tilde{u }_N(A,t)\) together with the spatial dependence of the coupling \(A(\mathbf R;d)\). A particularly simple analytic form is obtained in the motional-averaging regime, where the local bath relaxation rate is large compared with the relevant system--bath couplings, \(\gamma_{e1}\gg |A(\mathbf R;d)|\). In that limit the amplitude may be expanded as
\begin{equation}
\widetilde u_N(A,t)
=
1-A^2\,\kappa_N(t;\gamma_{e1})+\mathcal O(A^4),
\label{eq:single_spin_fastbath_uniform_revised}
\end{equation}
with \(\kappa_N\) real for the unpolarized, symmetric bath considered
here, so that
\begin{equation}
\ln\!\left\langle \mathcal L_N(t)\right\rangle
=
-\Gamma_v t
-\rho\,\kappa_N(t;\gamma_{e1})
\int d^2R\,A^2(\mathbf R;d)
+\mathcal O(A^4).
\label{eq:manyspin_fastbath_uniform_revised}
\end{equation}
For a dipolar-coupled two-dimensional surface bath, the coupling profile is
\begin{equation}
A(\mathbf R;d)=\frac{\mu_{0}\gamma_{e}^{2}\hbar}{4\pi r^{3}}\bigl(1-3\cos^{2}\theta\bigr),
\label{eq:appendix_dipolar_A}
\end{equation}
where \(\gamma_e\) denotes the electron gyromagnetic ratio, \(\mu_0\) is the vacuum permeability, \(r=\sqrt{R^2+d^2}\) is the separation between the NV center and the bath spin, and \(\theta\) is the angle between the NV axis and the vector joining the two spins. The geometric and dynamical content of the depth dependence must be kept separate. The profile obeys the scaling form
$A(\mathbf R;d)=C_{\mathrm{dd}}\,d^{-3}\,\mathfrak{a}(\mathbf R/d)$, where $C_{\mathrm{dd}}\equiv\mu_0\gamma_e^2\hbar/(4\pi)$ carries the dimensions and
$\mathfrak{a}(\mathbf x)=(1-3\cos^2\theta)/(1+|\mathbf x|^2)^{3/2}$ is dimensionless, since $\theta$ depends on $\mathbf R$ only through $\mathbf R/d$. The second spatial moment is therefore
\begin{equation}
\int d^2R\,A^2(\mathbf R;d)=\frac{C_{\mathrm{dd}}^2\,I_\mathfrak{a}}{d^4},
\qquad
I_a\equiv\int d^2x\,\mathfrak{a}^2(\mathbf x),
\label{eq:dipolar_depth_scaling_revised}
\end{equation}
where $I_\mathfrak{a}$ is a pure number fixed by the orientation of the NV axis relative to the surface normal. The $d^{-4}$ dependence is purely geometric: any second-order (weak-coupling) decoherence functional inherits it, and the bath dynamics enter only through the kernel $\kappa_N$, which has dimensions of time squared [Eq.~\eqref{eq:single_spin_fastbath_uniform_revised}]. Within the fast-bath expansion, Eq.~\eqref{eq:manyspin_fastbath_uniform_revised} becomes
\begin{equation}
\ln\langle\mathcal L_N(t)\rangle=-\Gamma_v t-\frac{\rho\,C_{\mathrm{dd}}^2\,I_\mathfrak{a}}{d^4}\,\kappa_N(t;\gamma_{e1})+O(A^4).
\label{eq:basic_scaling_uniform_revised}
\end{equation}
In the long-time motional-averaging regime, $\kappa_N(t;\gamma_{e1})\to t/(4\gamma_{e1})$, so the configurationally averaged coherence decays exponentially. Defining the coherence time through $\langle\mathcal L_N(t)\rangle=e^{-t/T_2}$ and using $I_\mathfrak{a}=3\pi/8$ for the NV axis along [111] of a (100) surface gives

\begin{equation}
\begin{split}
\frac{1}{T_{2}}
&=
\Gamma_v
+
\frac{\rho}{4\gamma_{e1}}
\int d^{2}R\,A(\mathbf R;d)^{2} \\
&=
\frac{\gamma_{v1}}{2}+\gamma_{v2}
+\frac{3\mu_{0}^{2}\gamma_{e}^{4}\hbar^{2}}{512\pi}\,
\frac{\rho}{\gamma_{e1}}\frac{1}{d^{4}},
\end{split}
\label{eq:T2_depth_appendix}
\end{equation}
the familiar \(T_2^{-1}\propto\rho/(\gamma_{e1}d^{4})\) depth law for a two-dimensional dipolar bath. If instead the decoherence functional grows as \(-\ln\mathcal L\propto d^{-4}t^{\,p}\) with \(p\neq1\), the same geometric coefficient produces \(T_2\propto d^{4/p}\): the exponent of the coherence-\emph{time} depth scaling encodes the dynamics, while the \(d^{-4}\) coefficient in \(-\ln\mathcal L\) encodes only the geometry.

Equations~\eqref{eq:dipolar_depth_scaling_revised}--\eqref{eq:T2_depth_appendix} should therefore be understood as asymptotic consequences of the motional-averaging limit, rather than as a universal depth law of the full noninteracting bath problem. An observed \(d^{-4}\)-type depth dependence alone does not uniquely prove motional narrowing.

\section{Central spin coupled to an interacting dissipative bath}
\label{meanfieldapproximation}

Let the two central-spin states used as the qubit have physical spin projections \(\nu_a=\langle a|\hat S_z^{(v)}|a\rangle\), \(a=0,1\), and write
\begin{equation}
\hat S_z^{(v)}\big|_{\mathcal Q}
=\bar \nu \,\hat I_v+\frac{\delta \nu}{2}\hat\sigma_z^{(v)},
\qquad
\bar\nu=\frac{\nu_0+\nu_1}{2},\quad
\delta \nu=\nu_0-\nu_1.
\label{eq:central_projection_mapping}
\end{equation}
For bath spins \(\hat S_z^{(i)}=\hat\sigma_z^{(i)}/2\), the physical longitudinal coupling \(\hat S_z^{(v)}\sum_i A_i\hat S_z^{(i)}\) becomes, in the toggling frame,
\begin{equation}
\hat H_N(t)=\hat H_B+
\frac{\bar \nu}{2}\sum_iA_i\hat\sigma_z^{(i)}+
\frac{\delta \nu}{4}y_N(t)\hat\sigma_z^{(v)}
\sum_iA_i\hat\sigma_z^{(i)}.
\label{eq:many_body_hamiltonian}
\end{equation}
The branch-independent term proportional to \(\bar \nu\) has no effect on the coherence for a noninteracting, initially diagonal bath, but it changes the detuning of interacting bath-spin pairs and must then be retained. For the NV \(m_s=0,-1\) qubit, \((\nu_0,\nu_1)=(0,-1)\), hence \(\bar \nu=-1/2\) and \(\delta \nu=1\); Eq.~\eqref{eq:many_body_hamiltonian} then contains the symmetric toggling term used in the main-text single-spin problem together with the required static bath field.

The secular bath Hamiltonian is
\begin{equation}
\hat H_B=
\sum_{i<j}J_{ij}\left(
\hat\sigma_+^{(i)}\hat\sigma_-^{(j)}+
\hat\sigma_-^{(i)}\hat\sigma_+^{(j)}\right)
+\sum_{i<j}U_{ij}\hat\sigma_z^{(i)}\hat\sigma_z^{(j)},
\label{eq:bath_hamiltonian_general}
\end{equation}

with \(\hat\sigma_+^{(i)}=|\!\uparrow_i\rangle\langle\downarrow_i\!|=(\hat\sigma_-^{(i)})^\dagger\) and \(\hat\sigma_z^{(i)}=|\!\uparrow_i\rangle\langle\uparrow_i\!|-|\!\downarrow_i\rangle\langle\downarrow_i\!|\). 
{The simulations use point-dipolar couplings. For two spin-\(1/2\) bath spins quantized along the applied field, the secular part of the dipole--dipole interaction is
\begin{equation}
\begin{aligned}
\hat H_{\mathrm{dd}}^{(ij)}
= {}& b_{ij}\bigl(1-3\cos^{2}\theta_{ij}\bigr)\\
&\times
\Bigl[\hat S_z^{(i)}\hat S_z^{(j)}
-\tfrac14\bigl(\hat S_+^{(i)}\hat S_-^{(j)}+\hat S_-^{(i)}\hat S_+^{(j)}\bigr)\Bigr],
\\
b_{ij}={}&\frac{\mu_0\gamma_e^{2}\hbar}{4\pi r_{ij}^{3}},
\end{aligned}
\label{eq:secular_dipolar_bath}
\end{equation}
where \(r_{ij}\) is the distance between the two spins and \(\theta_{ij}\) the angle between \(\mathbf r_{ij}\) and the field direction. With \(\hat S_z=\hat\sigma_z/2\) and \(\hat S_\pm=\hat\sigma_\pm\), Eq.~\eqref{eq:secular_dipolar_bath} takes the form of Eq.~\eqref{eq:bath_hamiltonian_general} with
\begin{equation}
\begin{aligned}
J_{ij}&=-\frac{b_{ij}}{4}\bigl(1-3\cos^{2}\theta_{ij}\bigr),\\
U_{ij}&=\frac{b_{ij}}{4}\bigl(1-3\cos^{2}\theta_{ij}\bigr)=-J_{ij}.
\end{aligned}
\label{eq:dipolar_JU}
\end{equation}
The longitudinal coupling \(U_{ij}\) multiplies \(\hat\sigma_z^{(i)}\hat\sigma_z^{(j)}=4\hat S_z^{(i)}\hat S_z^{(j)}\); the coefficient of \(\hat S_z^{(i)}\hat S_z^{(j)}\) is \(4U_{ij}=b_{ij}(1-3\cos^2\theta_{ij})\), the same normalization as the probe coupling \(A_i\) in Eq.~\eqref{eq:appendix_dipolar_A}.}
\par\medskip\noindent In the detuning test of Fig.~\ref{fig:validitymap}(d) and for the P1 bath of Sec.~\ref{P1_cent}, we add on-site energies to the bath Hamiltonian, $\hat H_B\to\hat H_B+\frac12\sum_i\varepsilon_i\hat\sigma_z^{(i)}$. In Fig.~\ref{fig:validitymap}(d), $\varepsilon_i$ is drawn independently from a Gaussian of zero mean and width $\sigma_\varepsilon$; for the P1 bath, $\varepsilon_i$ is the secular $^{14}$N hyperfine shift of center $i$, set by its Jahn--Teller axis and nuclear-spin projection \cite{park2022decoherence}. In both cases the on-site energies shift every pair detuning, $\Delta_{ij}^{(a)}\to\Delta_{ij}^{(a)}+\varepsilon_i-\varepsilon_j$, and leave the couplings unchanged.

The coherence is defined from the off-diagonal block
$\hat\tau_N(t)=\bra{0_v}\hat\rho_N(t)\ket{1_v}$,
\begin{equation}
\begin{aligned}
u_N(t)&=
\frac{\Tr_B[\hat\tau_N(t)]}{\Tr_B[\hat\tau_N(0)]},\\
\mathcal L_N(t)&=\operatorname{Re}\,u_N(t)
=e^{-\Gamma_vt}\operatorname{Re}\,\widetilde u_N(t).
\end{aligned}
\label{eq:many_body_coherence}
\end{equation}
We first consider \(\hat H_B=0\), for which the branch-independent field in Eq.~\eqref{eq:many_body_hamiltonian} does not change the population switching of any bath spin and the exact single-spin factors derived above apply independently.

\section{Pair projection, kinetic network, and local-rate approximation}
\label{sec:pair_projection_sm}

We now derive the rate used in the interacting-bath reduction. The bath master equation includes symmetric population relaxation and pure dephasing,
\begin{equation}
\begin{aligned}
\dot\rho={}&-i[\hat H_N(t),\rho]\\
&+\sum_\ell\frac{\gamma_{e1}}{2}
\left({\cal D}[\hat\sigma_+^{(\ell)}]+{\cal D}[\hat\sigma_-^{(\ell)}]\right)\rho
\\
&+\sum_\ell\frac{\gamma_{e2}}{2}{\cal D}[\hat\sigma_z^{(\ell)}]\rho,
\end{aligned}
\label{eq:full_master_compact}
\end{equation}
where \({\cal D}[\hat L]\rho=\hat L\rho\hat L^\dagger-\{\hat L^\dagger\hat L,\rho\}/2\), and \(\hat H_N\) is Eq.~\eqref{eq:many_body_hamiltonian}. The population imbalance of an isolated bath spin decays at \(\gamma_{e1}\), while its transverse coherence decays at \(\gamma_{e1}/2+\gamma_{e2}\).

For a pair \((i,j)\), use the ordered basis
\(
\{|a\rangle,|1\rangle,|2\rangle,|b\rangle\}
=\{|\uparrow_i\uparrow_j\rangle,|\uparrow_i\downarrow_j\rangle,
|\downarrow_i\uparrow_j\rangle,|\downarrow_i\downarrow_j\rangle\}
\).
In the flip-flop subspace \({\cal P}_{ij}=\mathrm{span}\{|1\rangle,|2\rangle\}\), define
\(\hat Z=|1\rangle\langle1|-|2\rangle\langle2|\) and
\(\hat X=|1\rangle\langle2|+|2\rangle\langle1|\). On central-spin branch \(a\), with physical projection \(\nu_a\), and for spectator eigenvalues \(s_k=\pm1\), the projected Hamiltonian is
\begin{equation}
\begin{aligned}
\hat H_{ij}^{(a)}&=J_{ij}\hat X+
\frac{\Delta_{ij}^{(a)}(\mathbf s)}{2}\hat Z,
\\
\Delta_{ij}^{(a)}(\mathbf s)&=
\nu_a(A_i-A_j)+2\sum_{k\ne i,j}s_k(U_{ik}-U_{jk}).
\end{aligned}
\label{eq:Delta_config_compact}
\end{equation}

The first term in \(\Delta_{ij}^{(a)}\) includes both the static and toggled pieces in Eq.~\eqref{eq:many_body_hamiltonian}; for the NV branches it is zero for \(m_s=0\) and \(-(A_i-A_j)\) for \(m_s=-1\). The spectator prefactor is two because the two flip-flop states have opposite \(\sigma_z^{(i)}\) and \(\sigma_z^{(j)}\), each spectator term \(U_{ik}s_k\hat\sigma_z^{(i)}\) contributing \(\pm U_{ik}s_k\) to the energies of \(|1\rangle\) and \(|2\rangle\). For the dipolar bath, Eq.~\eqref{eq:dipolar_JU} gives \(2\sum_{k}s_k(U_{ik}-U_{jk})=-2\sum_{k}s_k(J_{ik}-J_{jk})\), so the spectator shift is set by the same couplings that drive the flip-flops. The direct term \(U_{ij}\hat\sigma_z^{(i)}\hat\sigma_z^{(j)}\) shifts \(|1\rangle\) and \(|2\rangle\) equally and drops out of the detuning.

Let \(p_1=\rho_{11}\), \(p_2=\rho_{22}\), \(z=p_1-p_2\), and \(c=\rho_{12}\). The Hamiltonian gives
\(\dot z|_H=-4J_{ij}\operatorname{Im}c\) and
\(\dot c|_H=-i\Delta_{ij}^{(a)}c+iJ_{ij}z\).
Symmetric spin flips give \(\dot z|_1=-\gamma_{e1}z\) and damp \(c\) at \(\gamma_{e1}\). Pure dephasing leaves the populations unchanged. Since the bra and ket of \(c=\langle\uparrow_i\downarrow_j|\rho|\downarrow_i\uparrow_j\rangle\) differ on both spins, each channel \((\gamma_{e2}/2){\cal D}[\sigma_z]\) contributes \(-\gamma_{e2}c\), giving the total pair-coherence linewidth
\begin{equation}
\Gamma_{ij}=\gamma_{e1}+2\gamma_{e2}.
\label{eq:pair_Gamma_compact}
\end{equation}
The exact projected equations are therefore
\begin{equation}
\dot z=-\gamma_{e1}z-4J_{ij}\operatorname{Im}c,
\qquad
\dot c=-\left(\Gamma_{ij}+i\Delta_{ij}^{(a)}\right)c+iJ_{ij}z.
\label{eq:zc_pair_compact}
\end{equation}

The pair coherence cannot, in general, be eliminated adiabatically from Eq.~\eqref{eq:zc_pair_compact}. The population difference itself decays at \(\gamma_{e1}\), which is comparable to the coherence linewidth \(\Gamma_{ij}\), so setting \(\dot c\simeq0\) at fixed \(z\) is not justified by a separation of time scales. Factoring out the common decay, \(z=e^{-\gamma_{e1}t}\tilde z\) and \(c=e^{-\gamma_{e1}t}\tilde c\), Eq.~\eqref{eq:zc_pair_compact} becomes, exactly,
\begin{equation}
\dot{\tilde z}=-4J_{ij}\operatorname{Im}\tilde c,
\qquad
\dot{\tilde c}=-\left(\Gamma_{ij}-\gamma_{e1}+i\Delta_{ij}^{(a)}\right)\tilde c+iJ_{ij}\tilde z,
\label{eq:zc_pair_comoving}
\end{equation}
so that, relative to the population difference, the coherence relaxes only at \(\Gamma_{ij}-\gamma_{e1}=2\gamma_{e2}\). A Markovian transfer rate, linear in time, therefore exists only for \(|J_{ij}|\ll\sqrt{[\Delta_{ij}^{(a)}]^2+4\gamma_{e2}^2}\), where it equals \(4J_{ij}^2\gamma_{e2}/([\Delta_{ij}^{(a)}]^2+4\gamma_{e2}^2)\). For \(\gamma_{e2}=0\) no such rate exists, and Eq.~\eqref{eq:zc_pair_comoving} gives
\begin{equation}
z(t)=e^{-\gamma_{e1}t}\left[1-\frac{8J_{ij}^2}{\Omega_{ij}^2}\sin^2\!\frac{\Omega_{ij}t}{2}\right],
\label{eq:z_pair_gamma2_zero}
\end{equation}
which reduces to \(e^{-\gamma_{e1}t}\cos2J_{ij}t\) at \(\Delta_{ij}^{(a)}=0\), where $\Omega_{ij}=\sqrt{[\Delta_{ij}^{(a)}]^2+4J_{ij}^2}$.  

We therefore define the transfer rate through the zero-frequency response of the pair rather than through a Markov approximation. For \(z(0)=1\) and \(c(0)=0\), the Laplace transform \(\hat z(s)=\int_0^\infty e^{-st}z(t)\,dt\) of Eq.~\eqref{eq:zc_pair_compact} is
\begin{equation}
\hat z(s)=\left[s+\gamma_{e1}+\frac{4J_{ij}^2\,(s+\Gamma_{ij})}{(s+\Gamma_{ij})^2+[\Delta_{ij}^{(a)}]^2}\right]^{-1},
\label{eq:z_laplace_compact}
\end{equation}
and its value at \(s=0\), the time-integrated population difference \(\int_0^\infty z\,dt\), is reproduced exactly, for the given branch and spectator configuration and any \(J_{ij}\), by the exponential \(z(t)=e^{-(\gamma_{e1}+2W_{ij}^{(a)})t}\). Writing \(m_i=\langle\sigma_z^{(i)}\rangle\) and \(m_j=\langle\sigma_z^{(j)}\rangle\), one has \(z=(m_i-m_j)/2\). The exchange contribution becomes
\begin{equation}
\begin{aligned}
\dot m_i\big|_{ij}&=-W_{ij}^{(a)}(\mathbf s)(m_i-m_j),\\
W_{ij}^{(a)}(\mathbf s)&=
\frac{2J_{ij}^2\Gamma_{ij}}
{[\Delta_{ij}^{(a)}(\mathbf s)]^2+\Gamma_{ij}^2},
\end{aligned}
\label{eq:Wij_config_compact}
\end{equation}
with the opposite transfer for \(m_j\). Thus \(W_{ij}^{(a)}\) is a hopping rate, not an additional loss channel: it conserves \(m_i+m_j\) and relaxes the relative polarization \(m_i-m_j\) at rate \(2W_{ij}^{(a)}\), in addition to the intrinsic \(\gamma_{e1}\).

The rate \(W_{ij}^{(a)}\) preserves the correlation time \(\int_0^\infty z\,dt\) of the relative polarization, which controls the dephasing of the NV center by the pair in the motional-narrowing regime, and for \(\gamma_{e2}\gg\gamma_{e1}\) it reduces to the Markov rate above, which is therefore the same rate in that limit and not a second one. It does not reproduce the time dependence of Eq.~\eqref{eq:z_pair_gamma2_zero}: pointwise, the exact exchange contribution and the rate contribution to \(z(t)\) are both of order \(\lambda_{ij}^2=J_{ij}^2/([\Delta_{ij}^{(a)}]^2+\Gamma_{ij}^2)\), so the error of the rate description is of the same order. The condition
\begin{equation}
|J_{ij}|\ll\sqrt{[\Delta_{ij}^{(a)}(\mathbf s)]^2+\Gamma_{ij}^2}
\label{eq:weak_pair_condition}
\end{equation}
therefore states that the exchange of the pair is weak within its lifetime, so that treating it as a rate introduces an error of order \(\lambda_{ij}^2\); it does not imply that the exchange becomes a Markovian rate. Pairs that violate it are propagated explicitly (Sec.~\ref{sm:strong-group}), and the size of the error for the pairs that satisfy it is shown in Fig.~\ref{fig:Pair}.

To obtain a scalar rate for the local fluctuator model, we average over spectator configurations and over the central-spin populations,
\begin{equation}
\overline W_{ij}=
\sum_{a=0,1}p_a\left\langle
\frac{2J_{ij}^2\Gamma_{ij}}
{[\Delta_{ij}^{(a)}(\mathbf s)]^2+\Gamma_{ij}^2}
\right\rangle_{\mathbf s}.
\label{eq:Wij_average_compact}
\end{equation}
Here \(p_a\) is the population of branch \(a\); for the transverse initial state and symmetric central-spin relaxation used in the calculations, \(p_0=p_1=1/2\). If the central populations are unequal, the corresponding weights must be used. Unlike the symmetric pseudospin expression, Eq.~\eqref{eq:Wij_average_compact} does not assume that the two physical NV branches have opposite detunings. 

\par\medskip\noindent
\emph{Two-branch reduction of the coherence block.} Equations~\eqref{eq:zc_pair_compact}--\eqref{eq:Wij_average_compact} eliminate the pair on one branch at a time and then average the resulting rates. The measured object, however, is the off-diagonal central-spin block \(\hat\tau=\langle0|\hat\rho|1\rangle\), which evolves bilaterally: branch 0 acts from the left and branch 1 from the right. The same elimination can be carried out directly in this block, and doing so removes the branch average as a separate approximation. In the flip-flop subspace write \(p=\langle1|\hat\tau|1\rangle\), \(r=\langle2|\hat\tau|2\rangle\), \(c=\langle1|\hat\tau|2\rangle\), and \(\bar c=\langle2|\hat\tau|1\rangle\). With \(\hat H^{(0)}_{ij}\) on the left, \(\hat H^{(1)}_{ij}\) on the right, the linewidth of Eq.~\eqref{eq:pair_Gamma_compact}, and a branch-common phase factored out (it belongs to the global phase functional),
\begin{equation}
\begin{aligned}
\dot p&=-i\,\delta\!\Delta_{ij}\,p-iJ_{ij}(\bar c-c),\\
\dot r&=+i\,\delta\!\Delta_{ij}\,r-iJ_{ij}(c-\bar c),\\
\dot c&=-\big(\Gamma_{ij}+i\overline\Delta_{ij}\big)c-iJ_{ij}(r-p),\\
\dot{\bar c}&=-\big(\Gamma_{ij}-i\overline\Delta_{ij}\big)\bar c-iJ_{ij}(p-r),
\end{aligned}
\label{eq:twobranch_pair_eom}
\end{equation}
with
\begin{equation}
\begin{aligned}
\overline\Delta_{ij}(\mathbf s)&=\tfrac12\big[\Delta^{(0)}_{ij}+\Delta^{(1)}_{ij}\big]
\\
&=\bar \nu(A_i-A_j)+2\!\sum_{k\ne i,j}\!s_k(U_{ik}-U_{jk}),\\
\delta\!\Delta_{ij}&=\tfrac12\big[\Delta^{(0)}_{ij}-\Delta^{(1)}_{ij}\big]
=\tfrac{\delta \nu}{2}(A_i-A_j).
\end{aligned}
\label{eq:mean_detuning_def}
\end{equation}
Here, the intrinsic relaxation terms on $p$ and $r$ are omitted because they are common to both prescriptions.
The two cross-coherences precess at the branch-\emph{mean} detuning \(\overline\Delta_{ij}\); the branch \emph{difference} appears only as the phase \(\pm i\,\delta\!\Delta_{ij}\) of the diagonal components, which is exactly the pair's contribution to the hyperfine phase functional that the kinetic description retains without approximation. Under the weak-pair condition evaluated at \(\overline\Delta_{ij}\), the same zero-frequency reduction gives \(c\simeq-iJ_{ij}(r-p)/(\Gamma_{ij}+i\overline\Delta_{ij})\) and \(\bar c\simeq+iJ_{ij}(r-p)/(\Gamma_{ij}-i\overline\Delta_{ij})\), hence
\begin{equation}
\begin{aligned}
\dot p&=-i\,\delta\!\Delta_{ij}\,p+W_{ij}(\overline\Delta_{ij})\,(r-p),\\
\dot r&=+i\,\delta\!\Delta_{ij}\,r+W_{ij}(\overline\Delta_{ij})\,(p-r),
\end{aligned}
\label{eq:twobranch_rate_result}
\end{equation}
with \(W_{ij}(\overline\Delta)=2J_{ij}^2\Gamma_{ij}/(\overline\Delta^2+\Gamma_{ij}^2)\). The dispersive parts of the two eliminated coherences cancel in the combination \(\bar c-c\), so no Lamb-type shift survives at this order. The two-branch reduction therefore returns the \emph{same} kinetic form as the population-sector derivation, with one substitution: the transfer rate is evaluated at the branch-mean detuning rather than averaged over branch-resolved rates as in Eq.~\eqref{eq:Wij_average_compact}.

\begin{figure}
  \includegraphics[width=\columnwidth]{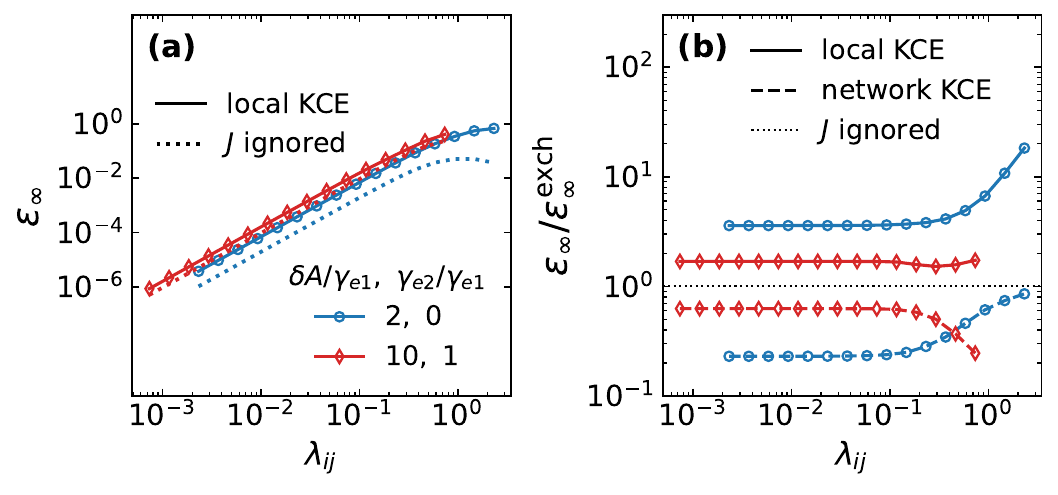}
\caption{
Comparison of pair approximations under Hahn echo.
(a) Maximum absolute coherence deviation from exact pair
propagation for the local-rate approximation (solid) and
neglecting the pair coupling (dotted).
(b) Deviations normalized by
$\epsilon_\infty^{\mathrm{exch}}
=\max_t|\mathcal L_{\mathrm{exact}}(J,t)
-\mathcal L_{\mathrm{exact}}(0,t)|$:
local rates (solid), correlated kinetic transfer (dashed),
and neglected pair coupling (dotted).
Colors identify the values of
$\delta A/\gamma_{e1}$ and $\gamma_{e2}/\gamma_{e1}$
shown in the legend.
}
  \label{fig:Pair}
\end{figure}

Two consequences follow. First, the scalar rate is sequence independent provided the pair's polarization difference, including the phase it acquires from the
probe and its reversal at each pulse, varies slowly on the scale
$1/\sqrt{\overline\Delta_{ij}^2+\Gamma_{ij}^2}$: a control \(\pi\) pulse exchanges the physical branches, \(\nu_0\leftrightarrow \nu_1\), which leaves \(\bar \nu\), and hence \(\overline\Delta_{ij}\) and \(W_{ij}(\overline\Delta_{ij})\), invariant; \(\delta\!\Delta_{ij}\) changes sign together with the toggling function \(y_N(t)\) and is carried exactly by the phase functional. Second, the two prescriptions do not bound each other. For a fixed spectator configuration, the branch-averaged rate is the larger when one branch is near resonance and the other strongly detuned, but for $\Delta^{(0)}_{ij}=-\Delta^{(1)}_{ij}=\delta$ the mean detuning is resonant and the branch-averaged rate is smaller by the factor $(1+\delta^2/\Gamma_{ij}^2)^{-1}$. We use the branch-averaged classifier throughout; the two prescriptions are not compared directly, and its adequacy rests on the agreement with ME-CCE4 (Figs. \ref{fig:validitymap}).

\par\medskip\noindent For many bath spins, pair elimination gives the kinetic network
\begin{equation}
\dot m_i=-\gamma_{e1}m_i-
\sum_{j\ne i}\overline W_{ij}(m_i-m_j).
\label{eq:many_spin_rate_compact}
\end{equation}
This is the controlled outcome of the weak-pair approximation. It retains the conservation law of exchange, polarization return, pair anticorrelations, and collective transport modes. A further, independent approximation is required to recover the product of single-spin fluctuators used in the main text. If neighboring spins remain close to an unpolarized local reservoir, \(m_j\simeq0\), then
\begin{equation}
\dot m_i\simeq-\gamma_i^{\rm eff}m_i,
\qquad
\gamma_i^{\rm eff}=\gamma_{e1}+\sum_{j\ne i}\overline W_{ij}.
\label{eq:gamma_eff_compact}
\end{equation}
The local closure discards the off-diagonal structure of the kinetic matrix. In particular, it no longer conserves total polarization under exchange, and it removes return processes and slowly relaxing diffusion modes. These are distinct errors from the pair-rate reduction in Eq.~\eqref{eq:weak_pair_condition}. Figure \ref{fig:Pair} shows the comparison of Eq.~\eqref{eq:many_spin_rate_compact} against Eq.~\eqref{eq:gamma_eff_compact}  and how the error accumulates when the correlated flips are ignored as in Eq.~\eqref{eq:gamma_eff_compact}. The local closure can also overestimate the exchange contribution of an isolated
pair: a flip-flop changes the probe field only through $\delta A=A_i-A_j$, whereas the local closure flips each spin independently, so for pairs with similar probe couplings the local-closure error can exceed the exchange contribution itself (Fig.~\ref{fig:Pair}(b)). 

{All calculations in this work use the branch-averaged rate of
Eq.~\eqref{eq:Wij_average_compact}; the two-branch form is given to show that the scalar rate is sequence independent and to make the difference
between the two prescriptions explicit.}

\subsection{Strong-group approximation}\label{sm:strong-group}

The rate description also fails for individual pairs for which Eq.~\eqref{eq:weak_pair_condition} is not satisfied. A configuration-averaged measure of the expansion parameter is
\begin{equation}
\lambda_{ij}^2=
\sum_{a=0,1}p_a\left\langle
\frac{J_{ij}^2}
{[\Delta_{ij}^{(a)}(\mathbf s)]^2+\Gamma_{ij}^2}
\right\rangle_{\mathbf s}
=\frac{\overline W_{ij}}{2\Gamma_{ij}}.
\label{eq:lambda_compact}
\end{equation}
The divergence of the golden-rule rate for a resonant pair as \(\Gamma_{ij}\to0\) is not physical; it signals that the rate reduction has been applied outside the domain of Eq.~\eqref{eq:weak_pair_condition}, where the pair instead undergoes coherent oscillations.

We choose a cutoff \(\lambda_c\) and call \((i,j)\) a strong link when \(\lambda_{ij}>\lambda_c\).  Two spins belong to the same group if they are
connected by a chain of strong links, which partitions the bath into disjoint
connected components \(\mathcal C=\{C_1,\ldots,C_M\}\). Singletons remain local fluctuators, while groups with two or more spins are propagated explicitly. To avoid double counting, internal links of a group are removed from the local rates,
\begin{equation}
\gamma_i^{\rm eff}=\gamma_{e1}+
\sum_{j\notin C(i)}\overline W_{ij}.
\label{eq:gamma_eff_hybrid}
\end{equation}

For each group,
\begin{equation}
\widetilde\ell_{C_\alpha}(t)=
\begin{cases}
\widetilde u_N(A_i,t;\gamma_i^{\rm eff}),&C_\alpha=\{i\},\\[3pt]
\widetilde\ell_{C_\alpha}^{\rm grp}(t;\{\gamma_i^{\rm eff}\}),&|C_\alpha|\ge2,
\end{cases}
\label{eq:group_factor_compact}
\end{equation}
where \(\widetilde u_N\) is the exact single-spin amplitude and
\(\widetilde\ell_{C_\alpha}^{\rm grp}=e^{\Gamma_v t}\,
\Tr_{C_\alpha}[\hat\tau_{C_\alpha}(t)]/\Tr_{C_\alpha}[\hat\tau_{C_\alpha}(0)]\)
is obtained by solving the full coherence-block master equation for all
spins in \(C_\alpha\), including every internal \(J_{ij}\) and
\(U_{ij}\), together with the intrinsic dephasing channels. The
superscript ``grp'' denotes the full finite-group propagation, which is
exact within the chosen cluster. Intergroup weak links enter only
through Eq.~\eqref{eq:gamma_eff_hybrid}. The factorized KCE
coherence is
\begin{equation}
\mathcal L_{\rm KCE}^{\rm loc}(t)
=e^{-\Gamma_v t}
\operatorname{Re}\!\left[\prod_{\alpha=1}^M\widetilde\ell_{C_\alpha}(t)\right],
\label{eq:L_hybrid_compact}
\end{equation}
This final product is the local independence assumption at the group level. It is accurate when coherent correlations remain confined within finite components and is expected to lose accuracy when moderately strong links form an extended network.

\section{Stochastic embedding of the correlated kinetic network}
\label{app:hsle}

This appendix specifies the network-resolved dynamics of
Sec.~\ref{sec:hsle}. Let $\mathcal C=\{\mathcal C_\alpha\}$ denote the
multispin quantum clusters and $\mathcal N$ the remaining singleton
spins. At fixed $\lambda_c$, flip-flop couplings within a cluster are
propagated quantum mechanically. Each rate-treated pair in distinct
clusters instead exchanges polarization through a joint event at rate
$\overline W_{ij}$. The rate is obtained from the pair reduction in
Appendix~\ref{sec:pair_projection_sm}; using the averaged rate for
probe coherence is the additional approximation discussed there.

\paragraph*{Singleton dynamics.}
Starting with $\mathbf s=(s_k)_{k\in\mathcal N}$, where $s_k=\pm1$, let
$F_k\mathbf s$ denote the configuration with $s_k$ flipped. For
symmetric intrinsic relaxation, the classical generator $K$ acts on
the configuration-conditioned coherence blocks as

\begin{equation}
\begin{aligned}
\sum_{\mathbf s'}K_{\mathbf s\mathbf s'}\hat\tau_{\mathbf s'}
={}&\frac{\gamma_{e1}}{2}
\sum_{k\in\mathcal N}
\bigl(\hat\tau_{F_k\mathbf s}-\hat\tau_{\mathbf s}\bigr)\\
&+\sum_{\substack{k<l\\k,l\in\mathcal N}}
\overline W_{kl}\frac{1-s_ks_l}{2}
\bigl(\hat\tau_{F_kF_l\mathbf s}-\hat\tau_{\mathbf s}\bigr).
\end{aligned}
\label{eq:K_appendix}
\end{equation}

The first line flips an individual singleton at rate
$\gamma_{e1}/2$. The second exchanges two singletons only when their
$z$ projections are opposite. Taking the first moment of
Eq.~\eqref{eq:K_appendix} gives their contributions to
Eq.~\eqref{eq:kinetic_network_main}.

\paragraph*{Conditioned cluster dynamics.}
The probe-coherence block
$\hat\tau_{\mathbf s}
=\langle0_v,\mathbf s|\hat\rho|1_v,\mathbf s\rangle$
is an operator on the multispin clusters. Its evolution is
Eq.~\eqref{eq:hsle_main}, with

\begin{equation}
\begin{aligned}
\mathcal K_{\mathcal C}[\mathbf s,t]\hat\tau
={}&-\Gamma_v\hat\tau
-i\hat H_{\mathcal C}^{(0)}(\mathbf s,t)\hat\tau
\\
&\quad
+i\hat\tau\hat H_{\mathcal C}^{(1)}(\mathbf s,t)
+\mathcal R_{\mathcal C}\hat\tau,\\
\hat H_{\mathcal C}^{(a)}(\mathbf s,t)
={}&\hat H_{B,\mathcal C}
+\sum_{i\in\cup_\alpha\mathcal C_\alpha}
\\
&\quad\times\Bigl[
\frac{A_i}{2}
\left(\bar \nu+(-1)^a\frac{\delta \nu}{2}y_N(t)\right)
\\
&\qquad
+\sum_{k\in\mathcal N}U_{ik}s_k
\Bigr]\hat\sigma_z^{(i)},\\
\Phi(\mathbf s)
={}&\frac{\delta \nu}{2}\sum_{k\in\mathcal N}A_ks_k ,
\\
\bar \nu={}&\frac{\nu_0+\nu_1}{2},
\qquad\delta \nu=\nu_0-\nu_1 .
\end{aligned}
\label{eq:hsle_terms}
\end{equation}

Here $\hat H_{B,\mathcal C}=\sum_\alpha\hat H_B^{(\alpha)}$, where $\hat H_B^{(\alpha)}$ contains the couplings $J_{ij}$ and $U_{ij}$ between spins of the same cluster $\mathcal C_\alpha$ [Eq.~\eqref{eq:dipolar_JU} restricted to $\mathcal C_\alpha$]; couplings between spins of different clusters enter only through $\overline W_{ij}$ and the spectator-averaged detunings of Eq.~\eqref{eq:pair_rate_main}. The term $-\Gamma_v\hat\tau$ includes the intrinsic probe-coherence decay exactly once. Intrinsic relaxation and pure dephasing of the cluster
spins are contained in

\begin{equation}
\begin{aligned}
\mathcal R_{\mathcal C}\hat\tau
=\sum_{i\in\cup_\alpha\mathcal C_\alpha}
\Bigg\{&
\frac{\gamma_{e1}}{2}\sum_{\eta=\pm}
\left[
\hat\sigma_\eta^{(i)}\hat\tau
\hat\sigma_\eta^{(i)\dagger}
-\frac12
\bigl\{
\hat\sigma_\eta^{(i)\dagger}
\hat\sigma_\eta^{(i)},\hat\tau
\bigr\}
\right]\\
&+\frac{\gamma_{e2}}{2}
\left[
\hat\sigma_z^{(i)}\hat\tau\hat\sigma_z^{(i)}
-\hat\tau
\right]\Bigg\}.
\end{aligned}
\label{eq:hsle_intrinsic}
\end{equation}

\paragraph*{Transfers involving clusters.}
The term $\mathcal T_{\mathbf s}$ in
Eq.~\eqref{eq:hsle_main} has two contributions,
$\mathcal T_{\mathbf s}
=\mathcal T_{\mathbf s}^{\mathcal C\mathcal N}
+\mathcal T_{\mathbf s}^{\mathcal C\mathcal C}$.
For a cluster spin $i$ and singleton $k$, define
$\hat\sigma_{s_k}^{(i)}=\hat\sigma_+^{(i)}$ when $s_k=+1$
and $\hat\sigma_-^{(i)}$ when $s_k=-1$, and let
$\hat P_{-s_k}^{(i)}$ project onto the cluster-spin state
opposite to $s_k$. Then

\begin{equation}
\begin{aligned}
\mathcal T_{\mathbf s}^{\mathcal C\mathcal N}
=\sum_{\alpha}\sum_{i\in\mathcal C_\alpha}
\sum_{k\in\mathcal N}\overline W_{ik}
\Big[
&\hat\sigma_{-s_k}^{(i)}
 \hat\tau_{F_k\mathbf s}
 \hat\sigma_{-s_k}^{(i)\dagger}\\
&-\frac12
 \bigl\{\hat P_{-s_k}^{(i)},
        \hat\tau_{\mathbf s}\bigr\}
\Big].
\end{aligned}
\label{eq:hsle_CN}
\end{equation}

The first term brings a transfer into configuration $\mathbf s$
from $F_k\mathbf s$; the second removes transfers out of
$\mathbf s$. Thus a singleton flip and a cluster-spin flip occur
in the same event.

A rate-treated pair joining two multispin clusters does not change
$\mathbf s$. With
$\hat B_{ij}^{(\eta)}=\hat\sigma_\eta^{(i)}\hat\sigma_{-\eta}^{(j)}$,
its \textit{formal} contribution is
\begin{equation}
\begin{aligned}
\mathcal T_{\mathbf s}^{\mathcal C\mathcal C}
={}&
\sum_{\alpha<\beta}
\sum_{\substack{i\in\mathcal C_\alpha\\ j\in\mathcal C_\beta}}
\overline W_{ij}\sum_{\eta=\pm}
\Big[
\hat B_{ij}^{(\eta)}\hat\tau_{\mathbf s}\hat B_{ij}^{(\eta)\dagger}
-\frac12\bigl\{\hat B_{ij}^{(\eta)\dagger}\hat B_{ij}^{(\eta)},
\hat\tau_{\mathbf s}\bigr\}\Big].
\end{aligned}
\label{eq:hsle_CC}
\end{equation}
It acts on both cluster blocks jointly, so the two clusters cannot be
propagated separately once such a link is retained. Both transfer
terms flip their two participating spins in opposite directions, and
in the absence of coherent intracluster exchange their first moments
reproduce the corresponding rate terms of
Eq.~\eqref{eq:kinetic_network_main}. Summing over $\mathbf s$ and
tracing over the clusters yields the probe coherence through
Eq.~\eqref{eq:hsle_coherence_main}. When there are no multispin
clusters, the transfer terms vanish and the construction reduces to
the classical jump network generated by $K$.

\paragraph*{Numerical implementation.}
The equations above define the model with every rate-treated
transfer retained. Our numerical implementation realizes them as follows.

\emph{Link classes.} A second threshold $0\le\lambda_w\le\lambda_c$
divides the rate-treated pairs. Pairs with $\lambda_{ij}\le\lambda_w$
are closed locally: their rates are added to the longitudinal
relaxation of both spins,
\begin{equation}
\gamma_i^{\rm eff}(\lambda_w)
=\gamma_{e1}
+\sum_{\substack{j\notin\mathcal C(i)\\ \lambda_{ij}\le\lambda_w}}
\overline W_{ij},
\label{eq:gamma_eff_hybrid_compact}
\end{equation}
where $\mathcal C(i)$ is the cluster containing $i$ (with
$\mathcal C(i)=\{i\}$ for a singleton), and $\gamma_i^{\rm eff}/2$
replaces $\gamma_{e1}/2$ in the single-spin flip terms of $K$ and of
$\mathcal R_{\mathcal C}$. Pairs with
$\lambda_w<\lambda_{ij}\le\lambda_c$ are retained as joint events.
$\lambda_w=0$ retains every rate-treated transfer permitted by the cluster-size and partner limits, and
$\lambda_w\ge\lambda_c$ recovers the local closure of
Sec.~\ref{sec:closure} exactly.

\emph{Cluster--cluster links.} A retained link between two multispin
clusters is not sampled. Taking these links in order of decreasing $\lambda_{ij}$, if the union of the two clusters does not exceed the cluster size limit, they are merged into one explicit
cluster and the link becomes an exactly propagated internal flip-flop;
otherwise the link is closed locally through
Eq.~\eqref{eq:gamma_eff_hybrid_compact}. Thus
$\mathcal T_{\mathbf s}^{\mathcal C\mathcal C}$ is realized either
exactly or in the local closure, never stochastically, and every
cluster block is propagated on its own. The cluster generator is diagonalized separately for each configuration of its singleton partners, so we limit each cluster to four partners: it keeps sampled links to the four singletons with the largest \(\lambda_{ij}\), with a single link to each, and all remaining cluster--singleton links are closed locally.

\emph{Sampling.}  Singletons with at least one retained link, to a cluster or to another singleton, are the live bits $\mathcal M\subseteq\mathcal N$, with configuration $\mathbf s=(s_k)_{k\in\mathcal M}$, so that every partner set satisfies $\mathcal P_\alpha\subseteq\mathcal M$. The remaining singletons and the multispin clusters with $\mathcal P_\alpha=\varnothing$ are not sampled: each is propagated on its own, a singleton through Eq.~\eqref{eq:reduced_2x2_system} with $\gamma_{e1}\to\gamma_k^{\mathrm{eff}}$, and their amplitudes $\widetilde\ell_{\mathcal C_\alpha}$ of Eq.~\eqref{eq:Lhybrid_main} multiply into the deterministic factor $D(t)$, with $D(t)=1$ if there are none.
Singleton flips and singleton--singleton exchanges
are sampled as the Markov jumps of $K$ with the exact
configuration-dependent rates. A retained cluster--singleton link
$(i,k)$ is unraveled as a quantum jump with unit trajectory weight:
between events each cluster block evolves under
Eq.~\eqref{eq:hsle_terms} plus the no-jump term
$-\tfrac12\overline W_{ik}\{\hat P_{-s_k}^{(i)},\cdot\}$ of
Eq.~\eqref{eq:hsle_CN}, which depends on the current value of $s_k$;
events are proposed at the constant rate $\overline W_{ik}$ and
accepted with probability
\begin{equation}
\pi_{ik,n}=
\frac{\operatorname{Tr}[\hat P_{-s_k}^{(i)}\widetilde\varrho_{\alpha,n}]}
{\operatorname{Tr}\widetilde\varrho_{\alpha,n}},
\label{eq:hsle_accept}
\end{equation}
where $\widetilde\varrho_{\alpha,n}$ is the population block of the
cluster along trajectory $n$. An accepted event applies
$\hat\sigma_{s_k}^{(i)}\cdot\hat\sigma_{s_k}^{(i)\dagger}$, with $s_k$ the value before the event, to the coherence and population blocks of the cluster and then flips $s_k$.
The population block is propagated with
the \textit{branch-mean generator}, the average of the two probe-branch
Liouvillians in Eq.~\eqref{eq:hsle_terms}, so that one sampled event
sequence serves every pulse sequence; the coherence block itself is
propagated with the two branch-resolved generators, exchanged at each
$\pi$ pulse. 

\textit{Fields.} Within an explicit cluster all couplings $J_{ij}$ and $U_{ij}$ are retained exactly. In the implementation, the singleton sums in Eq.~\eqref{eq:hsle_terms} run over live bits only. The probe field $\Phi(\mathbf s)=(\delta\nu/2)\sum_{k\in\mathcal M}A_k s_k$ is retained exactly along each trajectory, and a spin $i\in\mathcal C_\alpha$ sees the Ising field $\sum_{k\in\mathcal P_\alpha}U_{ik}s_k$ of the live bits $\mathcal P_\alpha\subset\mathcal M$ linked to its cluster. A link-free singleton acts on the probe through its amplitude in $D(t)$, which accounts exactly for its own field. Its Ising field on cluster spins, like the fields from spins in other clusters, is not propagated and enters only through the spectator-averaged detunings in $\overline W_{ij}$ and $\lambda_{ij}$, Eq.~\eqref{eq:pair_rate_main}; neglecting this back-action is what allows the link-free singletons to factorize out of the trajectory average.

\paragraph*{Trajectory estimator.}
Trajectory $n$ carries the live-bit configuration $\mathbf s_n(t)$, the
coherence and population blocks $\widetilde\tau_{\alpha,n}(t)$ and
$\widetilde\varrho_{\alpha,n}(t)$ of every cluster $\alpha$ that owns a
retained link, both started from the same unpolarized initial block
and propagated without the intrinsic probe decay,
$\widetilde\tau_{\alpha,n}=e^{\Gamma_vt}\tau_{\alpha,n}$, and the phase
$\phi_n(t)=\int_0^t du\,y_N(u)\Phi(\mathbf s_n(u))$. Let
\begin{equation}
X_n(t)=e^{-i\phi_n(t)}\prod_{\alpha}
\frac{\operatorname{Tr}\widetilde\tau_{\alpha,n}(t)}
     {\operatorname{Tr}\widetilde\varrho_{\alpha,n}(t)},
\label{eq:hsle_Xn}
\end{equation}

which satisfies $X_n(0)=1$; the intrinsic probe decay cancels in each ratio. With $D(t)$ the deterministic factor defined above, the coherence estimate is
\begin{equation}
\widehat{\mathcal L}_{\mathrm{KCE}}^{\mathrm{net}}(t)
=e^{-\Gamma_vt}\operatorname{Re}\!\left[
  D(t)\frac{1}{N_{\mathrm{traj}}}
  \sum_{n=1}^{N_{\mathrm{traj}}}X_n(t)
\right],
\label{eq:net_KCE}
\end{equation}
Equation~\eqref{eq:net_KCE} estimates Eq.~\eqref{eq:hsle_coherence_main} for the implemented generator, Eq.~\eqref{eq:hsle_main} with the link classes, cluster merging, partner limit, and field treatment described above, and it does so without bias. Each population block evolves under a trace-preserving generator plus the no-jump term, so $\prod_\alpha\operatorname{Tr}\widetilde\varrho_{\alpha,n}(t)$ is the probability density of the sampled event record, apart from the singleton events, whose rates are exact. The coherence blocks follow the same record, and expanding Eq.~\eqref{eq:hsle_main} in these events gives $\sum_{\mathbf s}\operatorname{Tr}\hat\tau_{\mathbf s}(t)$ as the sum over records of $e^{-\Gamma_vt}D(t)e^{-i\phi_n(t)}\prod_\alpha\operatorname{Tr}\widetilde\tau_{\alpha,n}(t)$. The trace ratios are therefore exact likelihood weights, and the branch-mean population generator affects only the variance. The statistical error is $e^{-\Gamma_vt}$ times the standard error of the mean of $\operatorname{Re}[D(t)X_n(t)]$; all calculations here use $\Gamma_v=0$. All results use
$\lambda_w=0.1$, clusters of at most four spins, at most four partners
per cluster, and $4000$ trajectories ($8000$ at
$\rho=0.001$~nm$^{-2}$); the convergence in $\lambda_w$ and
$N_{\mathrm{traj}}$ is shown in Appendix~\ref{benchmarks}.

\section{Configuration average for a two-dimensional random bath}
\label{C-average}
We consider a two-dimensional bath of \(n\) spins distributed over a
surface region \(S\) of area \(|S|\), with areal density
\begin{equation*}
\rho=\frac{n}{|S|}.
\end{equation*}
We assume that the bath-spin positions \(\mathbf R_i\) are independently
and uniformly distributed within \(S\), as appropriate for a spatially
random surface bath. In the noninteracting limit, the coherence
factorizes into independent single-spin amplitudes,
Eq.~\eqref{eq:factorized_coherence_uniform},
\begin{equation}
\mathcal L_N\!\left(t\,\middle|\,\{\mathbf R_i\}\right)
=
e^{-\Gamma_v t}
\operatorname{Re}\!\left[\prod_{i=1}^{n}
\widetilde u_N(A(\mathbf R_i;d),t)\right],
\label{eq:appendix_config_factorized}
\end{equation}
where \(A(\mathbf R_i;d)\) is the coupling between the central spin at
depth \(d\) and bath spin \(i\), and \(\widetilde u_N(A,t)\) is the
single-spin amplitude of Eq.~\eqref{eq:reduced_2x2_system}.

Averaging Eq.~\eqref{eq:appendix_config_factorized} over independently
distributed bath-spin positions, the average commutes with the real part
and factorizes,
\begin{equation}
\begin{split}
\left\langle \mathcal L_N(t)\right\rangle
&=
e^{-\Gamma_v t}
\operatorname{Re}\prod_{i=1}^{n}
\left\langle
\widetilde u_N(A(\mathbf R_i;d),t)
\right\rangle \\
&=
e^{-\Gamma_v t}
\operatorname{Re}\left[
\frac{1}{|S|}
\int_S d^2R\,
\widetilde u_N(A(\mathbf R;d),t)
\right]^n .
\end{split}
\label{eq:appendix_config_avg_finite}
\end{equation}
Using \(n=\rho |S|\) and writing the spatial average as
\begin{equation}
\begin{aligned}
\frac{1}{|S|}\int_S d^2R\,
&\widetilde u_N(A(\mathbf R;d),t)
=
1+\frac{I_N(t)}{|S|},\\
I_N(t)&=\int_S d^2R\,\Big[\widetilde u_N\big(A(\mathbf R;d),t\big)-1\Big],
\end{aligned}
\end{equation}
Eq.~\eqref{eq:appendix_config_avg_finite} becomes
\begin{equation}
\left\langle \mathcal L_N(t)\right\rangle
=e^{-\Gamma_v t}\operatorname{Re}\left[1+\frac{I_N(t)}{|S|}\right]^{\rho |S|}.
\label{eq:appendix_config_avg_before_limit}
\end{equation}
Taking the thermodynamic limit \(|S|\to\infty\) at fixed areal density
\(\rho\), we obtain
\begin{equation}
\left\langle \mathcal L_N(t)\right\rangle
=
e^{-\Gamma_v t}
\operatorname{Re}\exp\left[
\rho\int d^2R\,
\Big(
\widetilde u_N(A(\mathbf R;d),t)-1
\Big)
\right],
\label{eq:appendix_config_avg_final}
\end{equation}
which reproduces Eq.~\eqref{eq:config_average_uniform}. For an
unpolarized bath with symmetric rates, \(\widetilde u_N\) is real and
the real part can be dropped. The derivation assumes that the bath-spin
positions are statistically independent and that the coherence
factorizes into independent single-spin amplitudes.

\section{Numerical benchmark results}
\label{benchmarks}

\emph{Benchmark protocol.} Bath realizations are random placements of about $200$ electron spins on the diamond surface; the NV is a spin-1 central spin ($D=2.88$~GHz, $B=30$~mT along the NV axis, $m_s=0/{-1}$ branch), and couplings are point-dipolar. Intrinsic bath relaxation enters as symmetric longitudinal relaxation with rate $\gamma_{e1}=1/T_{1e}$. All results are averaged over 100 bath realizations. ME-CCE averages each realization over 100 sampled pure bath states (except in the cutoff test of Fig.~10); KCE starts from the unpolarized bath state and samples only spectator configurations (Appendix~\ref{app:sampling}) and, for network KCE, transfer trajectories (Appendix~\ref{sec:hsleTrajConv}). An absolute validation against exact propagation of the full dissipative Liouvillian, at $T_{1e}=80~\mu s$, is given for a dilute 12-spin bath in the main text (Fig.~\ref{fig:validation}).  Figure~\ref{fig:validationSlow} repeats this validation at $T_{1e}=0.8$~ms, where $J_{\mathrm{typ}}/\gamma_{e1}=10.5$: ME-CCE4--6 agree with the exact result within their sampling error of 0.017, and network and local KCE deviate by at most 0.05 and 0.06.

\begin{figure}
  \includegraphics[width=0.7\columnwidth]{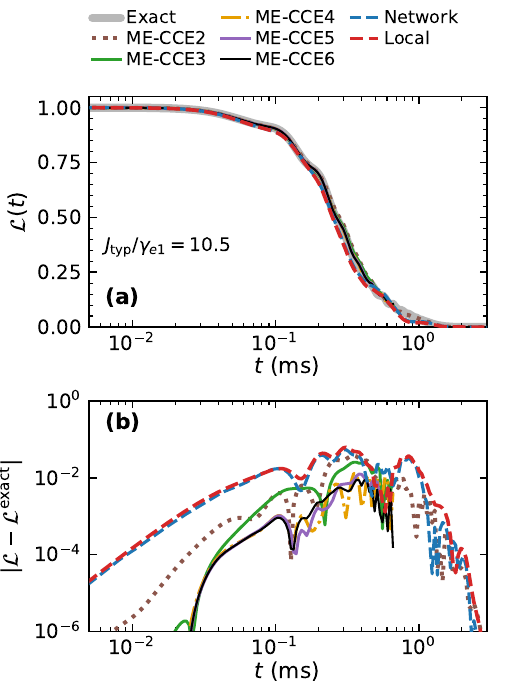}
\caption{Exact benchmark for the 12-spin bath of Fig.~\ref{fig:validation} at $T_{1e}=0.8$~ms, $J_{\mathrm{typ}}/\gamma_{e1}=10.5$ ($\rho=0.001~\mathrm{nm}^{-2}$, $d=10$~nm). (a)~Hahn-echo coherence from exact propagation of the full dissipative Liouvillian (gray), ME-CCE2--6 averaged over 100 sampled pure bath states, and network and local KCE ($\lambda_c=0.5$). (b)~Absolute pointwise error $|\mathcal L(t)-\mathcal L^{\mathrm{exact}}(t)|$. The sampling error of the ME-CCE average is at most 0.017; ME-CCE4--6 agree with the exact result within it, while network and local KCE deviate by at most 0.05 and 0.06.}
\label{fig:validationSlow}
\end{figure}

\textit{ME-CCE convergence.}
(Fig.~\ref{fig:MECCE4}) shows the convergence test of ME-CCE in the expansion order, up to $k=5$, at $d=d_{ee}$ for both surface densities and for the P1 bath, at $T_{1e}=20$ and $100~\mu$s. In the dilute surface bath all orders coincide, so intrabath correlations are negligible there. In the dense surface bath, ME-CCE4 and ME-CCE5 agree both at $T_{1e}=20~\mu$s and $100~\mu$s. We also show that for the P1 bath at $T_{1e}=100~\mu$s ME-CCE4 is also converged.

\begin{figure}
  \includegraphics[width=\columnwidth]{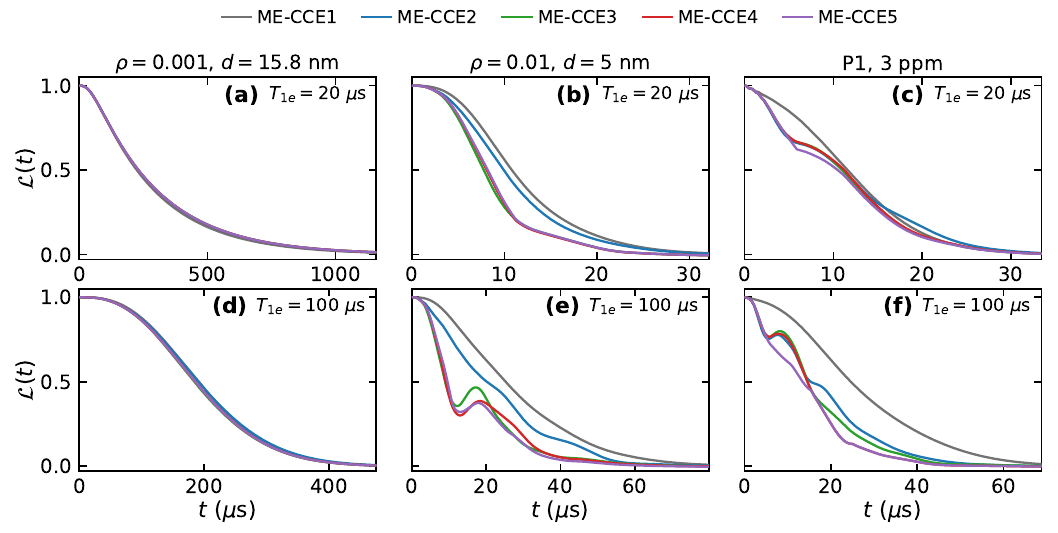}
\caption{Convergence of ME-CCE in the expansion order. Hahn-echo coherence from ME-CCE($k$), $k=1$--5, at $d=d_{ee}$ for surface baths with (a,d) $\rho=0.001~\mathrm{nm}^{-2}$ ($d=15.8$~nm) and (b,e) $\rho=0.01~\mathrm{nm}^{-2}$ ($d=5$~nm), and (c,f) for the three-dimensional P1 bath at 3~ppm, with $T_{1e}=20~\mu$s (top) and $100~\mu$s (bottom).}
  \label{fig:MECCE4}
\end{figure}

\emph{Bath-relaxation dependence.} Figure~\ref{fig:t1eSM} shows the corresponding sweep versus $T_{1e}$ at $d=10$~nm, i.e. at $d/d_{ee}=0.63$, $1.4$, and $2.0$ for the three densities. The maps in Fig.~\ref{fig:t1eSM}(d--f) show $\Delta\mathcal L(T_{1e},t)=\mathcal L_{\mathrm{KCE}}^{\mathrm{loc}}-\mathcal L_{\mathrm{MECCE4}}$ for local KCE at $\lambda_c=0.5$. For $T_{1e}\lesssim20~\mu$s the difference stays below $0.05$ at all three densities. For slower relaxation it grows within the decay window: at $\rho=0.01$ KCE decays faster, by up to $0.15$ at $T_{1e}=1.6$~ms, and at $\rho=0.005$ the difference changes sign near $T_{1e}=60~\mu$s and reaches $0.13$ at $T_{1e}=1.6$~ms. At $\rho=0.001$ it stays below $0.08$.

\begin{figure*}
  \includegraphics[width=0.85\textwidth] {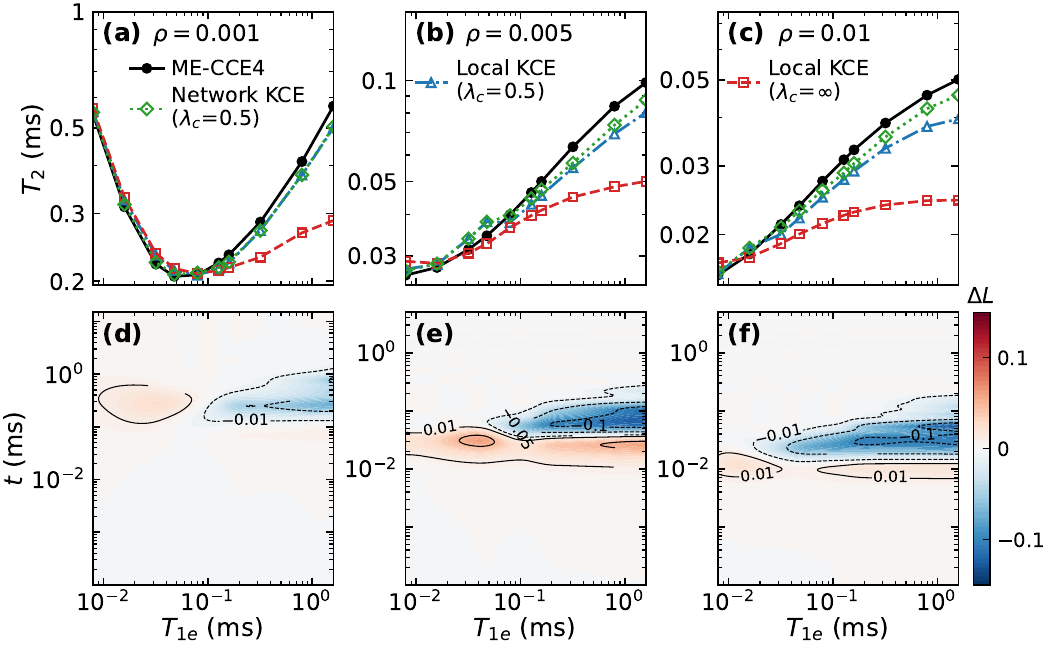}
 \caption{
Bath-relaxation dependence of Hahn-echo coherence at
$d=10$~nm. Columns correspond to
$\rho=0.001$, $0.005$, and $0.01~\mathrm{nm}^{-2}$.
(a--c) Coherence time versus $T_{1e}$ for ME-CCE4,
network KCE, local KCE at $\lambda_c=0.5$,
and the singleton local limit.
(d--f) Coherence signal differences
$\Delta\mathcal L(T_{1e},t)$ for the local KCE relative to ME-CCE4.
Negative values indicate lower local KCE coherence.
}
  \label{fig:t1eSM}
\end{figure*}

\emph{Cutoff convergence.} Figure~\ref{fig:lcconvSM} shows the $\lambda_c$ convergence of the maximum pointwise deviation $\epsilon_\infty$. For this convergence test only, we use an unsampled ME-CCE4 reference, with no bath-state sampling noise, and average over 100 bath realizations. At $\rho=0.001$~nm$^{-2}$, the deviation decreases monotonically as strong pairs are resolved ($\epsilon_\infty=0.034\to0.0015$ from $\lambda_c=\infty$ to $0.3$), whereas at the denser baths it saturates at a density-dependent floor that remains flat from $\lambda_c=1$ down to $0.3$, indicating that the residual deviation from the ME-CCE4 reference does not reside in unresolved strong pairs. 
Lowering $\lambda_c$ brings additional pairs into quantum clusters and reduces the deviation at low density, with little change at high density. We do not lower it below $\lambda_c\simeq0.3$: the strong-link graph then percolates, and before the four-spin limit is applied, typical realizations contain components of seven or eight spins. These components grow rapidly as $\lambda_c$ decreases toward the percolation threshold, beyond which the largest one scales with the bath size, so no fixed cluster size suffices; with the four-spin limit, the additional pairs in these components would instead be returned to transfer rates.

\emph{Weak-link threshold convergence.} Figure~\ref{fig:lwconvSM} shows the convergence of network KCE in $\lambda_w$ at $\rho=0.01~\mathrm{nm}^{-2}$ and $d=10$~nm for slow bath relaxation at $T_{1e}=0.32$ and $0.8$~ms, where the intercluster transfers matter most: the local closure ($\lambda_w=\lambda_c$) deviates from network KCE by up to $0.11$ and gives a $T_2$ mpre than $25\%$ shorter than ME-CCE4. Lowering $\lambda_w$ from $0.3$ to $0.01$ changes the coherence by at most $6\times10^{-3}$ and $T_2$ by less than $1\%$, so the value $\lambda_w=0.1$ used throughout is converged.

\begin{figure}
  \includegraphics[width=\columnwidth]{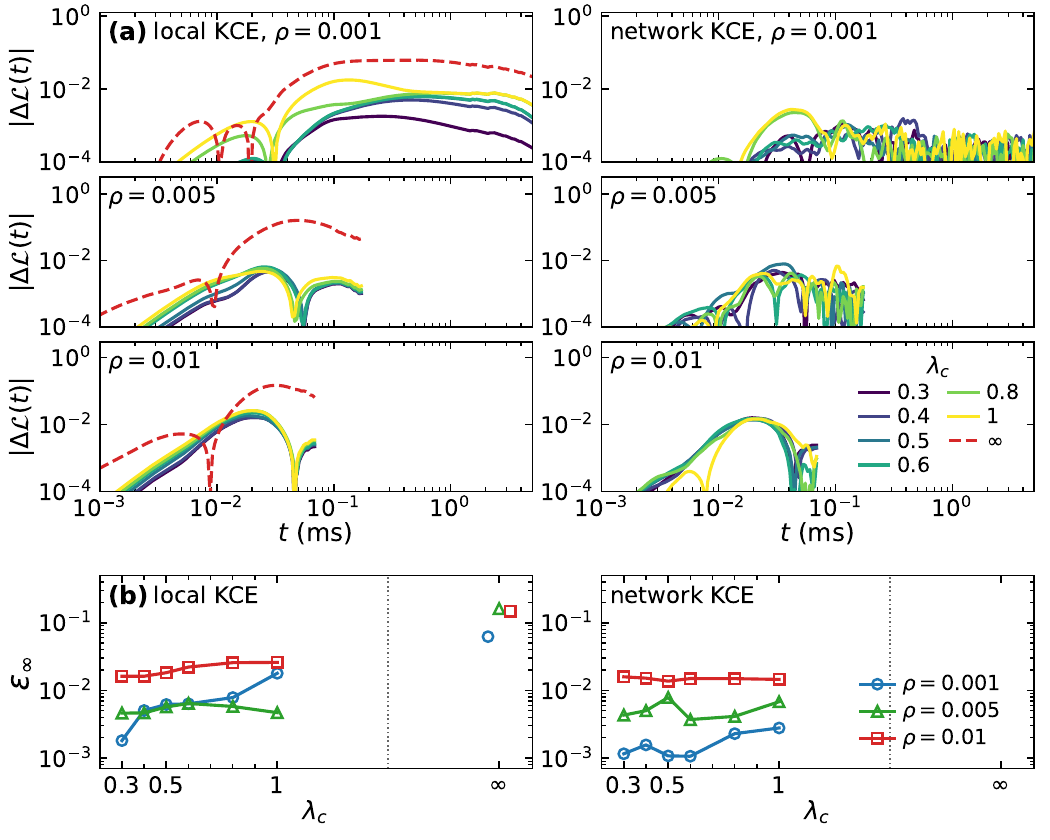}
  \caption{Cutoff dependence of local KCE (left) and network KCE
(right) at $d=10$~nm and $T_{1e}=20~\mu s$.
(a) shows $|\Delta\mathcal L(t)|$ relative to
ME-CCE4; rows correspond to
$\rho=0.001$, $0.005$, and $0.01~\mathrm{nm}^{-2}$.
Colors denote $\lambda_c$ as indicated.
(b) shows the maximum pointwise deviation
$\epsilon_\infty$ versus $\lambda_c$.}
  \label{fig:lcconvSM}
\end{figure}

\begin{figure}
  \includegraphics[width=\columnwidth]{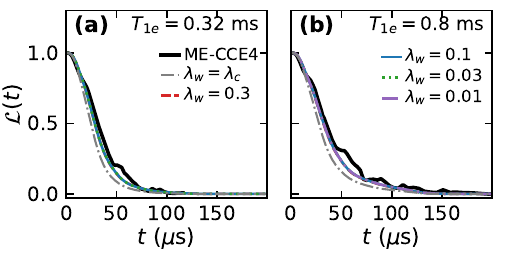}
  \caption{Convergence of network KCE in the weak-link threshold
$\lambda_w$ for 
$\rho=0.01~\mathrm{nm}^{-2}$ at $d=10$~nm, for (a)
$T_{1e}=0.32$~ms and (b) $T_{1e}=0.8$~ms (at $\lambda_c=0.5$, and 4000
trajectories). Network KCE at $\lambda_w=0.3$, $0.03$ and, in (b),
$0.01$ differs from $\lambda_w=0.1$ by at most $6\times10^{-3}$
and gives the same $T_2$ to within $1\%$. The local closure
($\lambda_w=\lambda_c$) and ME-CCE4, averaged over 100 bath
states, are shown for comparison.}
  \label{fig:lwconvSM}
\end{figure}

\subsection{Convergence of the spectator-configuration average}
\label{app:sampling}
The spectator polarizations shift the pair detuning in
Eq.~\eqref{eq:pair_hamiltonian_main}. Computing
$\overline W_{ij}$ and $\lambda_{ij}$ therefore requires an
average over their configurations and over the two probe states.
Here we retain the 12 spectators nearest to either spin of the pair
and sum over the probe states with weights $p_a$. For every KCE curve
of Fig.~\ref{fig:validation}, where the 12-spin bath leaves 10
spectators per pair, we average exactly over all $2^{10}$  specator configurations, and in all other calculations, we average exactly over all polarization configurations $2^{12}$. The state-resolved calculations of
Fig.~\ref{fig:validitymap}(d,e) take the spectator field of each
sampled bath state and need no average.

We checked both the number of spectators and the number of
sampled configurations. The selected 12 account for a median
$99.6\%$ of the summed squared coupling differences
$\sum_k(U_{ik}-U_{jk})^2$, which set the variance of the
spectator-induced detuning [Fig.~\ref{fig:spectators}(a)]. Retaining
16 or 20 spectators changes the KCE coherence by at most
$2.3\times10^{-4}$ across the three densities tested; retaining only
eight changes it by up to $2\times10^{-2}$ at
$\rho=0.01~\mathrm{nm}^{-2}$ [Fig.~\ref{fig:spectators}(b)].

Figure~\ref{fig:sampling} compares sampled averages to that of the exhaustive $2^{12}$ configurations at
$\rho=0.005~\mathrm{nm}^{-2}$. The median of the maximum
pointwise coherence change decreases approximately as
$N_{\mathrm{samp}}^{-1/2}$. At these conditions ($J_{\mathrm{typ}}/\gamma_{e1}=2.9$), 128 samples give a median deviation of $3\times10^{-4}$. For slower relaxation the Lorentzian in $\overline W_{ij}$ narrows and the average is dominated by rare near-resonant configurations: for $J_{\mathrm{typ}}/\gamma_{e1}\geq10$, 128 samples shift the KCE coherence by up to about $0.3$, which the exact average removes at negligible extra cost.

\begin{figure}
\centering
\includegraphics[width=\columnwidth]
{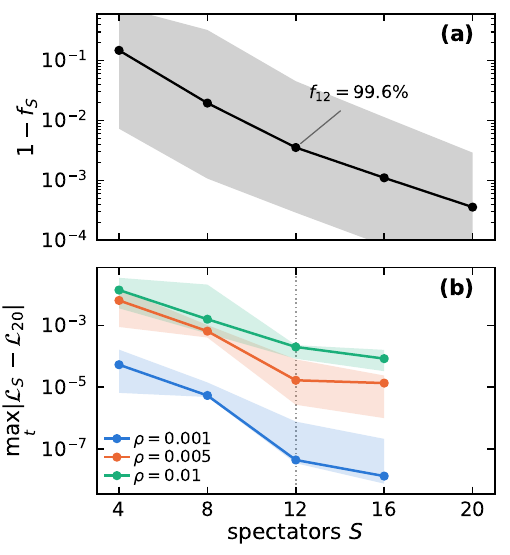}
\caption{Convergence in the number of spectators $S$ retained in the configuration average, at $d=10$~nm, $T_{1e}=100~\mu$s, under spin echo.
(a) Fraction $1-f_S$ of the spectator-detuning variance left out by the first $S$ spectators, where $f_S=\sum_{k\leq S}(U_{ik}-U_{jk})^2/\sum_k(U_{ik}-U_{jk})^2$ with spectators ordered as in the classifier; averaged (line) and range (shading) over 271 pairs of one bath realization, the same at all three densities.
(b) Maximum pointwise difference between the local KCE coherence with $S$ spectators and with $S=20$, for 100 bath realizations at each density. Median over 100 bath realizations (line) and their range (shading). The dotted line marks the $S=12$ used here.}
\label{fig:spectators}
\end{figure}

\begin{figure}
\centering
\includegraphics[width=\columnwidth]
{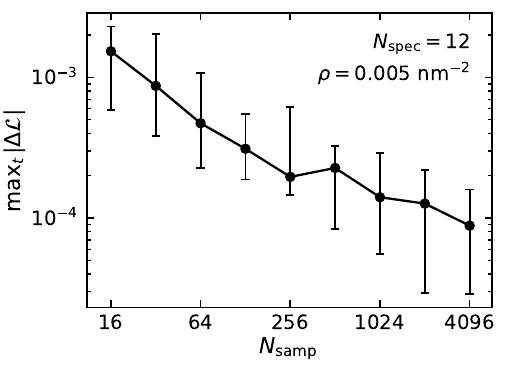}
\caption{Convergence of the sampled spectator-configuration average
at $\rho=0.005~\mathrm{nm}^{-2}$, $d=10$~nm, and
$T_{1e}=20~\mu s$ under spin echo, retaining 12 spectators.
We plot $\max_t|\Delta\mathcal L(t)|$, where
$\Delta\mathcal L=\mathcal L_{\mathrm{KCE}}^{(N_{\mathrm{samp}})}
-\mathcal L_{\mathrm{KCE}}^{\mathrm{exh}}$ and
$\mathcal L_{\mathrm{KCE}}^{\mathrm{exh}}$ uses an exhaustive
average over all $2^{12}$ configurations. Circles are medians
over three bath realizations with ten sampling seeds each;
bars show the interquartile range. The point at
$N_{\mathrm{samp}}=4096$ is still a sampled average, not the
exhaustive reference. 
}
\label{fig:sampling}
\end{figure}

\subsection{Trajectory convergence of network KCE}
\label{sec:hsleTrajConv}

For each bath realization, network KCE averages
$N_{\mathrm{traj}}$ independent trajectories. With $X_n(t)$
and $D(t)$ defined Eq~\eqref{eq:hsle_Xn} in the trajectory estimator above, we
estimate the standard error of the full coherence at each
time point as
\[
\delta\mathcal L(t)=
\left[
\frac{\widehat{\operatorname{Var}}_n
\bigl[\operatorname{Re}(D(t)X_n(t))\bigr]}
{N_{\mathrm{traj}}}
\right]^{1/2}.
\]
Figure~\ref{fig:hsleTrajConv} takes the largest value of
$\delta\mathcal L(t)$ on each echo trace, then reports the
median over 100 bath realizations. It falls from about
$2\times10^{-2}$ at $N_{\mathrm{traj}}=50$ to
$2.6\times10^{-3}$ at $3200$, consistent with
$N_{\mathrm{traj}}^{-1/2}$ scaling. For a fixed time grid, the
sampling cost grows approximately linearly with
$N_{\mathrm{traj}}$. The network-KCE calculations reported in the main text use $4000$ trajectories per realization ($8000$ at $\rho=0.001$~nm$^{-2}$), for which the median
standard error is about $2.3\times10^{-3}$.

\begin{figure}
\centering
\includegraphics[width=\columnwidth]{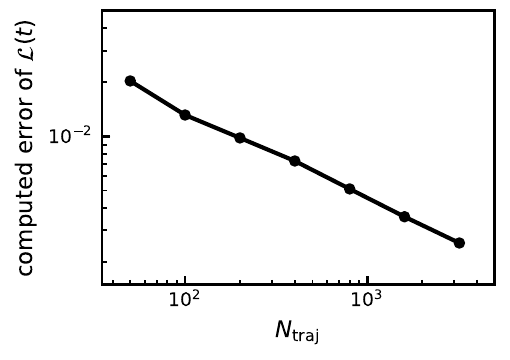}
\caption{
Trajectory convergence of network KCE at
$\rho=0.01~\mathrm{nm}^{-2}$, $d=10$~nm,
$T_{1e}=20~\mu s$, $\lambda_c=0.5$, and
$\lambda_w=0.1$ under spin echo. Each point is the
median over 100 bath realizations of the largest
estimated standard error of the coherence on the
time grid.
}
\label{fig:hsleTrajConv}
\end{figure}

\subsection{CPMG filters and noise spectra}
\label{app:cpmgspectra}
A sequence of $N$ $\pi$ pulses at $t_j=(2j-1)t/2N$ multiplies the bath field seen by the probe by a toggling function $y(s)=\pm1$. For Gaussian noise the decay $\chi(t)=-\ln\mathcal L(t)$ is the integral of the noise spectrum $S(\omega)$ weighted by the filter $|Y_N(\omega)|^2$, where $Y_N$ is the Fourier transform of $y$~\cite{biercuk2011dynamical,degen2017quantum}. For $N\geq2$ the filter is dominated by a peak at $\omega=\pi N/t$ whose width decreases as $1/t$. In this narrow-filter limit $\chi(t)\simeq(t/\pi)\,S(\pi N/t)$~\cite{alvarez2011measuring}, and each point of a CPMG decay with $0.1<\mathcal L<0.9$ gives an estimate of $S$ at $\omega=\pi N/t$.

Figure~\ref{fig:cpmgfilters}(d--f) applies this inversion to the CPMG decays computed at $d=10$~nm and $T_{1e}=100~\mu$s for $N=2$ to $64$, for one bath realization and every method. The inversion assumes Gaussian noise and is therefore approximate where a few strongly coupled spins dominate; since it is applied in the same way to all methods, it compares the noise that each method implies. At $\rho=0.001~\mathrm{nm}^{-2}$ the spectra of all methods agree to within about a third and fall close to $\omega^{-2}$, the Lorentzian tail that produces the $N^{2/3}$ growth of $T_2$~\cite{Lange2010}. At $\rho=0.005$ and $0.01~\mathrm{nm}^{-2}$ the methods separate. ME-CCE1 underestimates the noise by a factor of two or more at every frequency, because it omits the field fluctuations produced by flip-flops. The singleton-local limit overestimates it, most strongly at higher frequencies, because it replaces coherent pairs by fast switching rates that spread their noise over a broad band. Network and local KCE stay within about $40\%$ of ME-CCE4 over the frequency range. At $\rho=0.01~\mathrm{nm}^{-2}$, ME-CCE4 and both KCE formulations show a local maximum near $60$ to $80$~kHz, comparable to the typical flip-flop coupling at this density, which neither ME-CCE1 nor the singleton-local limit reproduces; we attribute it to coherent exchange within the explicitly treated clusters.

Figure~\ref{fig:cpmgfilters}(a--c) shows which part of each spectrum drives the decay. For $N=8$ and $t$ equal to the ME-CCE4 coherence time, the filter selects a narrow window around $\omega=\pi N/t$, and the area under $S(\omega)|Y_N(\omega)|^2$ sets the decay reached at that time. At $\rho=0.001~\mathrm{nm}^{-2}$ the weighted spectra of all methods agree to within about $20\%$. At the higher densities ME-CCE1 places too little noise in this window, so its coherence decays too slowly, whereas the singleton-local limit places too much, about twice the ME-CCE4 value at $\rho=0.01~\mathrm{nm}^{-2}$, so its coherence decays too fast. Network and local KCE place nearly the same noise in the window as ME-CCE4, consistent with their agreement in $T_2$.

\begin{figure}
\centering
\includegraphics[width=\columnwidth]
{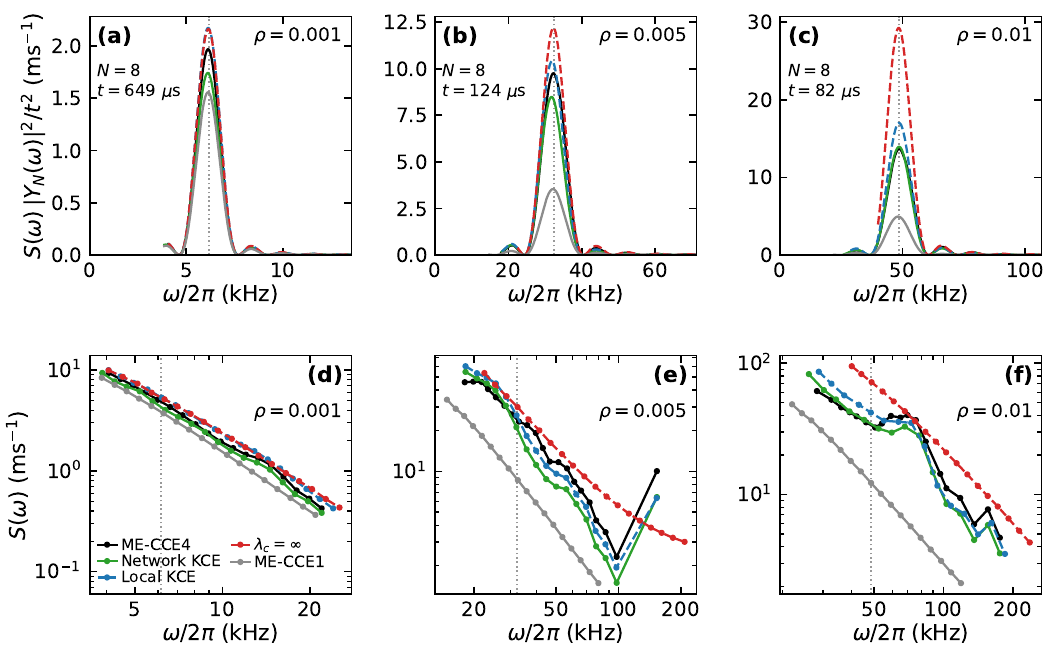}
\caption{Noise sampled by CPMG for each method. (a--c)~Noise spectrum weighted by the CPMG filter, $S(\omega)|Y_N(\omega)|^2/t^2$, for $N=8$ pulses at $t$ equal to the ME-CCE4 coherence time, for (a)~$\rho=0.001$, (b)~$0.005$, and (c)~$0.01~\mathrm{nm}^{-2}$. Here $Y_N$ is the Fourier transform of the toggling function of the sequence, and $S(\omega)$ of each method is interpolated from (d--f). (d--f)~Noise spectrum $S(\omega)$ obtained from the CPMG decays ($N=2$--$64$) using $S(\pi N/t)\simeq\pi\chi(t)/t$, with $\chi=-\ln\mathcal L$; the dotted line marks the filter peak $\omega=\pi N/t$ of (a--c). All results are for $d=10$~nm, $T_{1e}=100~\mu$s, and one bath realization: ME-CCE4 (100 sampled bath states), network and local KCE ($\lambda_c=0.5$), the singleton-local limit ($\lambda_c\to\infty$), and ME-CCE1.}
\label{fig:cpmgfilters}
\end{figure}

\section{Number of clusters and computational cost}
\label{sec:clusterCountsSM}
\begin{table*}
\caption{
KCE clusters and ME-CCE clusters for identical
bath realizations at $d=10$~nm and $T_{1e}=20~\mu s$ (mean $\pm$
standard deviation). Each spin belongs to one KCE cluster;
the size-weighted KCE counts therefore sum to $N$, up to rounding.
The ``above-cutoff pairs'' column counts pairs with
$\lambda_{ij}>\lambda_c$ before applying the four-spin size limit.
The number in parentheses counts above-cutoff pairs reassigned to
transfer rates because they cannot be retained within the allowed
clusters; it counts couplings, not spins or clusters. The ME-CCE columns enumerate connected candidate clusters within $r_{\mathrm{dipole}}=6\,d_{ee}$ ($95$, $42$, and $30$~nm at $\rho=0.001$, $0.005$, and $0.01~\mathrm{nm}^{-2}$); at a fixed number of spacings, all three densities give nearly the same counts, because the number of spins within that distance is the same.
}
\label{tab:clusterCountsSM}
\small
\setlength{\tabcolsep}{2.5pt}
\begin{ruledtabular}
\begin{tabular}{lccccccccc}
& & & \multicolumn{4}{c}{KCE clusters by size}
& \multicolumn{3}{c}{{ME-CCE clusters by order}} \\
\cline{4-7}\cline{8-10}
$\rho$ (nm$^{-2}$) & $N$ & \shortstack{above-cutoff pairs\\(capped)}
& 1 & 2 & 3 & 4 & 2 & 3 & 4 \\
\colrule
$0.001$ & $199\pm15$ & $38\pm6$ ($0$)
& $123\pm9$ & $36\pm7$ & $0.7\pm0.8$ & $0.25\pm0.45$
& $2.4(4)\times10^{3}$ & $3.4(1.1)\times10^{4}$
& $5.6(2.4)\times10^{5}$ \\
$0.005$ & $197\pm14$ & $83\pm9$ ($2.2\pm1.1$)
& $56\pm6$ & $55\pm7$ & $6.6\pm2.2$ & $3.1\pm1.9$
& $2.3(4)\times10^{3}$ & $3.4(1.0)\times10^{4}$
& $5.5(2.4)\times10^{5}$ \\
$0.01$ & $196\pm15$ & $103\pm10$ ($6.5\pm3.6$)
& $40\pm5$ & $51\pm6$ & $9.2\pm3.3$ & $6.6\pm2.9$
& $2.3(4)\times10^{3}$ & $3.4(1.0)\times10^{4}$
& $5.5(2.4)\times10^{5}$ \\
\end{tabular}
\end{ruledtabular}
\end{table*}

\begin{table*}
\caption{{
Computational cost at $\rho=0.01~\mathrm{nm}^{-2}$,
$N\simeq200$, $d=10$~nm, and $T_{1e}=20~\mu s$ for spin echo
at 101 time points. Wall times exclude program startup and
were measured on one 28-core node using 28 MPI ranks. Local and network KCE use $\lambda_c=0.5$
and clusters of at most four spins. All the ME-CCE calculations contain the clusters sizes listed in 
Table~\ref{tab:clusterCountsSM}. Network KCE uses
$\lambda_w=0.1$, 4000 trajectories per realization. Core-hours assume all 28 cores are allocated, and the last column gives sequential wall time for 100 realizations. 
}}
\label{tab:costSM}
\begin{ruledtabular}
\begin{tabular}{lccc}
Method & Wall time / realization & Core-hours / realization
& 100 realizations (serial) \\
\colrule
ME-CCE2
& $0.4$~s & $0.0034$ & $44$~s \\
Local KCE
& $1$~s & $0.008$ & $1.7$~min \\
ME-CCE3
& $15$~s & $0.12$ & $26$~min \\
Network KCE ($N_{\mathrm{traj}}=4000$)
& $6.5\times10^{2}$~s & $5.1$ & $18$~h \\
ME-CCE4 
& $4.6\times10^{3}$~s & $36$ & $5.4$~days \\
\end{tabular}
\end{ruledtabular}
\end{table*}

We compare cluster counts using identical bath realizations. At each
$\rho\in\{0.001,0.005,0.01\}~\mathrm{nm}^{-2}$, we generate
realizations with $N\simeq200$ spins, choosing the bath area to keep
$N$ approximately fixed. The bath radius therefore scales as
$\rho^{-1/2}$. We set $d=10$~nm and $T_{1e}=20~\mu s$ and construct the KCE
clusters with $\lambda_c=0.5$, and above-cutoff pairs are added in order of decreasing \(\lambda_{ij}\), and a pair that would create a cluster of more than four spins is skipped and assigned its polarization-transfer rate.

For the same realizations, we count the clusters at each ME-CCE order from 2 to 4 with \(r_\mathrm{dipole} = 6\,d_{ee}\)~\cite{onizhuk2024understanding}. Table~\ref{tab:clusterCountsSM} lists the number of clusters at each order, and Table~\ref{tab:costSM} the corresponding computation times. For ME-CCE1, the number of clusters equals the number of bath spins, \(N\).

The KCE partition contains, on average, approximately $160$, $121$,
and $107$ clusters at increasing density, of which $37$, $65$, and
$67$ are multispin clusters. Under the local closure, singleton
factors follow from the exact single-spin solution; network-resolved
KCE additionally samples the retained intercluster transfers.
The number of above-cutoff pairs rises from $38$ to $103$ across the
three densities. At the two higher densities, the four-spin limit
reassigns $2.2\pm1.1$ and $6.5\pm3.6$ of these pairs to rates,
respectively. By comparison, the ME-CCE graph at $\rho = 0.01~nm^{-2}$ contains about $5.5\times10^{5}$ connected four-spin candidates within $6\,d_{ee}$, compared with about 7 four-spin clusters in KCE. 

Cluster counts alone do not determine computational cost. In Table~\ref{tab:costSM} we report the measured wall times on one 28-core node. Local KCE takes 1~s per realization and ME-CCE2 $0.4$~s. ME-CCE3 with all $3.4\times10^{4}$ triples takes $15$~s. Sampling the retained intercluster transfers with 4000 trajectories raises the network-KCE time to $6.5\times10^{2}$~s. ME-CCE4 with all $5.5\times10^{5}$ quadruples takes $4.6\times10^{3}$~s per realization, about \(4.6\times10^{3}\) times longer than local KCE and 7 times longer than network KCE with 4000 sampled trajectories, while both KCE predictions of \(T_2\) at this condition differ from ME-CCE4 by less than 4\%.

\clearpage
\phantomsection
\label{sm:references}
\bibliographystyle{apsrev4-2}
\bibliography{Reference}

\end{document}